\documentclass[12pt,preprint]{aastex}

\usepackage{epsfig}

\newcommand{\Vlsr}{$V_\mathrm{LSR}$}

\newcommand{\variancemeritfII}{$\sigma( F_\mathrm{II}(B_\mathrm{i}))$}
\newcommand{\meritfIII}{$F_\mathrm{III}(\mathrm{B_i})$}    %% \times \sigma( F_\mathrm{II}(B_\mathrm{i}))$}
\newcommand{\meritfII}{$F_\mathrm{II}(\mathrm{B_i})$}
\newcommand{\meritfn}{$f_\mathrm{n}(B_\mathrm{i})$}  %%(\mathrm{B_\mathrm{best}})$}
\newcommand{\Gfact}{\emph{$G_\mathrm{n}$}}
\newcommand{\Vcliclsr}{V$_\mathrm{CLIC,LSR}$}
\newcommand{\Vliclsr}{V$_\mathrm{IBEX,LSR}$}
\newcommand{\Bnofil}{B$_\mathrm{POL}$}
\newcommand{\PAnofil}{$\theta_\mathrm{POL}$}
\newcommand{\PAfil}{PA$_\mathrm{filament}$}
\newcommand{\Bibex}{B$_\mathrm{IBEX}$}
\newcommand{\Bfil}{$B_\mathrm{FIL}$}
\newcommand{\Ball}{$B_\mathrm{ALL}$}

\newcommand{\microG}{$\mu$G}

\newcommand{\sintheta}{\emph{sin}($\theta_\mathrm{n,i}$)}
\newcommand{\thetan}{$\theta_\mathrm{n}$}
\newcommand{\PA}{$\theta_\mathrm{PA}$}
\newcommand{\PAgal}{$\theta_\mathrm{gal}$}
\newcommand{\PAra}{$\theta_\mathrm{RA}$}
\newcommand{\thirteen}{thirteen}
\newcommand{\seven}{seven}
\newcommand{\five}{six}
\newcommand{\dPA}{$\Delta \theta$}

\newcommand{\Pol}{$P$}
\newcommand{\dPol}{$\Delta P$}
\newcommand{\Polpct}{$P_\%$}
\newcommand{\Polpctpr}{$P^\prime _\%$}

\newcommand{\HeI}{He$^\circ$}

\newcommand{\gL}{$\ell$}
\newcommand{\gB}{$b$}

\def\PAcel{PA$_\mathrm{RA}$}

\def \deeg {$^\circ$}
\def \ebv {E(B-V)}

\def \kms {${\rm km~s}^{-1}$}
\def\cmtwo{cm$^{-2}$}
\def\cc{cm$^{-3}$}
\def\HH{H$_\mathrm{2}$}

\def\HI{H$^\mathrm{o}$}

\def\MgI{Mg$^\mathrm{o}$}

\def\NHI{N(H$^\mathrm{o}$)}
\newcommand{\NH}{N(H)}
\def\NHH{N(H$_\mathrm{2}$)}
\def\NHII{N(H$^\mathrm{+}$)}
\def\NH2{N(H$_\mathrm{2}$)}

\def\HII{H$^\mathrm{+}$}
\def\mG{$\mu$G}
\def\glon{$\ell$}
\def\elon{$\lambda$}
\def\elat{$\beta$}
\def\glat{$b$}

\slugcomment{Submitted to Astrophysical Journal June 24, 2015}

\shorttitle{Structure in the local interstellar magnetic field}
\shortauthors{Frisch et al.}

\begin{document}

\title{Charting the Interstellar Magnetic Field causing the Interstellar Boundary Explorer (IBEX) Ribbon of Energetic Neutral Atoms}

\author{P. C. Frisch}
\affil{Dept. Astronomy and Astrophysics, University of Chicago,
Chicago, IL  60637}
%\email{frisch@oddjob.uchicago.edu}
%\and
\author{A. Berdyugin and V. Piirola}
\affil{Finnish Centre for Astronomy with ESO, University of Turku, Finland}
\author{A. M. Magalhaes and D. B. Seriacopi}
\affil{ Inst. de Astronomia, Geofisica e Ciencias Atmosfericas, Universidade de Sao Paulo,
  Brazil}
\author{S. J.  Wiktorowicz}
\affil{Dept. Astronomy, University of California at Santa Cruz, Santa Cruz, CA}
\author{B-G Andersson}
\affil{SOFIA Science Center, USRA, Moffett Field, CA}
\author{H. O. Funsten}
\affil{Los Alamos National Laboratory, Los Alamos, NM}
\author{D. J. McComas\altaffilmark{1}}
\affil{Southwest Research Institute, San Antonio, TX}
\altaffiltext{1}{Also University of Texas, San Antonio, TX}
\author{N. A. Schwadron}
\affil{Space Science Center, University of New Hampshire}
\author{J. D. Slavin}
\affil{Harvard-Smithsonian Center for Astrophysics, 
  Cambridge, MA}
\author{A. J. Hanson}
\affil{School of Informatics and Computing, Indiana University, Bloomington, IN}
\author{C.-W. Fu}
\affil{School of Computer Engineering, Nanyang Technological University, Singapore}

\begin{abstract} The interstellar magnetic field (ISMF) near the
heliosphere is a fundamental component of the solar galactic
environment {that can only be studied using polarized starlight.  The
results of an ongoing survey of the linear polarizations of local
stars are analyzed with the goal of linking the interstellar magnetic
field that shapes the heliosphere to the nearby field in interstellar
space. We present new results on the direction of the magnetic field
within 40 pc obtained from analyzing polarization data using a merit
function that determines the field direction that provides the best
fit to the polarization data.}  Multiple magnetic components are
identified, including a dominant interstellar field, \Bnofil, that is
aligned with the direction \glon,\glat$= 36.2^\circ,49.0^\circ$ ($\pm
16.0^\circ$).  Stars tracing \Bnofil\ have the same mean distance as
stars that do not trace \Bnofil, but show weaker average polarizations
consistent with a smaller column density of polarizing material.
\Bnofil\ is aligned with the ISMF traced by the IBEX Ribbon to within
$7.6^{+14.9}_{-7.6}$ degrees.  The variations in the polarization
position angle directions derived from the data that best match
\Bnofil\ indicate a low level of magnetic turbulence, $\sim 9^\circ
\pm 1^\circ$.  The direction of \Bnofil\ is obtained after excluding
polarization data tracing a separate magnetic structure that appears
to be associated with interstellar dust deflected around the
heliosphere.  The velocities of local interstellar clouds relative to
the local standard of rest (LSR) increase with the angles between the
LSR velocities and \Bnofil, indicating that the kinematics of local
interstellar material is ordered by the ISMF.  The Loop I superbubble
that extends close to the Sun contains dust that reddens starlight and
whose distance is determined by the color excess \ebv\ of starlight.
Polarizations caused by grains aligned with respect to \Bnofil\ are
consistent with the location of the Sun in the rim of the Loop I
superbubble.  An angle of $76.8^{+23.5}_{-27.6}$ between \Bnofil\ and
the bulk LSR velocity the local interstellar material indicates a
geometry that is consistent with an expanding superbubble.  The
efficiency of grain alignment in the local ISM has been assessed using
stars where both polarization data and hydrogen column density data
are available. Nearby stars appear to have larger polarizations than
expected based on reddened sightlines, which is consistent with
previous results, but uncertainties are large.
%%%%%results from Fosalaba et al. (2002) indicating a nonlinear increase in
%%%%%linear polarization strengths with decreasing reddening.  
Optical polarization and color excess \ebv\ data indicate the presence of
nearby interstellar dust in the BICEP2 field.  Color excess \ebv\
indicates an optical extinction of $A_\mathrm{V} \sim 0.59 $ in the
BICEP2 field, while the polarization data indicate that $A_\mathrm{V}
> 0.09 $  mag.  The IBEX Ribbon ISMF extends to the boundaries of the
BICEP2 region.  \end{abstract}

\keywords{ISM: clouds, dust, magnetic fields --- Physical processes:  polarization ---Sun:  heliosphere}

%%%%%%%%%%%%%%%%%%%%%%%%%%%%%%%%%%%%%%%%%%%%%%%%%%%%%%%%%%%%%%%%%%%%%%%%%%%%%%%%%%%%%%%%%%%%%%%%%%%%%%%
\newpage
\section{Introduction}\label{sec:intro}

The Sun is traveling through a dynamically evolving interstellar
environment that contains low density, magnetized, partially ionized
interstellar material traveling rapidly away from the center of the
Loop I superbubble.  The heliosphere is shaped by the solar wind
interaction with the interstellar gas and magnetic field.  A unique
diagnostic of the direction of the magnetic field at the heliosphere
location is provided by the Ribbon of energetic neutral atoms (ENAs)
discovered by the Interstellar Boundary Explorer
\citep[IBEX,][]{McComas:2009sci,Schwadron:2009sci,Funsten:2009sci,Funsten:2013,Schwadron:2011sep}.
The flux of TeV galactic cosmic rays at the Earth is controlled by the
ISMF direction that is traced by the IBEX Ribbon
\citep{Schwadron:2014gcr}.  Similarly, galactic cosmic ray fluxes onto
exoplanets and their astrospheres (stellar wind bubbles) depend on the
interstellar magnetic field that surrounds the exoplanet system
\citep{Frisch:1993gstar}.  Understanding the role of the interstellar
magnetic field (ISMF) in the past and future galactic environments of
the solar system and nearby exoplanet systems requires knowledge of the
magnetic field configuration and its connection to the magnetized and
partially ionized medium in the galactic neighborhood of the
heliosphere.  The purpose of this study is to connect the ISMF that
shapes the heliosphere with the extended magnetic field in the nearby
interstellar clouds.

Studies over the past century of optical and ultraviolet (UV)
interstellar absorption lines, and the reddening of starlight by
interstellar dust, have revealed the physical properties, dynamics and
distribution of local interstellar material within 40 pc of the Sun
\citep{Frisch:2011araa}.  Although the Sun was known to be located in
a region of very low interstellar dust densities
\citep{Fitzgerald:1968}, measurements of polarized starlight proved to
be a viable method for testing the very local interstellar magnetic
field \citep{Piirola:1977,Tinbergen:1982}.  Nevertheless, the
interstellar dust content and magnetic field configuration close to
the heliosphere have been enigmatic.  In this paper we present new
high-sensitivity measurements of polarized starlight that allow
filling the gaps in our knowledge of the configuration of the local
interstellar magnetic field within 40 pc of the Sun.

Polarized starlight provides the only method for tracing the ISMF
direction in the low density interstellar medium (ISM) near the solar
system.  For polarization due to dichroic extinction, starlight
becomes linearly polarized while traversing a medium formed by
charged asymmetric submicron-sized interstellar dust grains that are
aligned with the most opaque grain axis oriented perpendicular to the
interstellar magnetic field
\citep{HoangLazarian:2014radiativetorques}.  Comparisons between
polarized dust emission, polarized synchrotron emission, and linearly
polarized optical starlight indicate that optical polarization
position angles are oriented parallel to the ISMF direction in the
nearby ISM \citep{Frischetal:2014icns}.  The low interstellar column
densities near the Sun
\citep{BohlinSavageDrake:1978,FrischYork:1983,Woodetal:2005} impose
two conditions on the study of the local ISMF: high-sensitivity
polarization measurements are required, and the analysis method must
utilize statistically-weighted data so that low-significance
polarization data can be included.

The goal of charting the direction of the ISMF within 40 pc is to
connect the interstellar field in the solar neighborhood with the
interstellar magnetic field that shapes the heliosphere. In
\citet[][Paper I]{Frisch:2010ismf1} we derived the orientation of the
local ISMF by assuming that the field has a dipole configuration, and
that the pole of this field could be retrieved by applying a
minimization algorithm to the polarization position angles.  The
analysis was based on polarization data in the literature.  New data
were collected on the polarizations of nearby stars in order to fully
the constrain the magnetic field direction \citep[][Paper
II]{Frisch:2012ismf2}.  Applying a minimization procedure with
statistically weighted position angles, to the extended data set
produced a interstellar field direction close to the direction of the
IBEX Ribbon ISMF.  The two directions were within $ \sim 33 ^\circ \pm
27^\circ $ of each other.  The velocity vector of the bulk flow of the
cluster of local interstellar clouds (CLIC) relative to the Local
Standard of Rest (LSR) was found to form an angle of $\sim 76^\circ$
with the best-fitting ISMF to the polarization data.  Subsequently, we
identified a distinct nearby magnetic structure that appears to be
formed by aligned dust grains entrained in the interstellar magnetic
field draping over the heliosphere \citep{Frisch:2015fil}.

The present study includes new data that have been collected in both
the northern and southern hemispheres in order to map the magnetic
field in nearby interstellar space, described in Section (\S)
\ref{sec:data}. {We refine the analysis method used in Papers I and II
where a merit function is used to establish the best-fitting ISMF
direction to the ensemble of interstellar polarization measurements
(\S \ref{sec:method}).  The best-fitting ISMF direction to all
qualifying polarization data is given in \S \ref{sec:best}.
Polarization data associated with a distinct filamentary polarization
structure are identified (\S \ref{sec:filament}).  Omission of the
filament stars from the data sample gives a best-fitting ISMF
direction in close agreement with the IBEX ISMF direction (\S
\ref{sec:domismf}) although not all significant polarizations agree
(\S \ref{sec:phasespace}).  }  Magnetic
turbulence is estimated from the dispersion of polarization position
angles (\S \ref{sec:turb}).  Additional possible unrecognized local
ISMF components are briefly considered (\S \ref{sec:otherismf}).  The
broad implications of these results are discussed in \S
\ref{sec:discussion}, including the relation between \Bnofil\ and the
IBEX Ribbon (\S \ref{sec:ibex}), galactic cosmic ray asymmetries (\S
\ref{sec:gcr}), the origin of the separate polarization filament (\S
\ref{sec:filamentdisc}), the gas-dust relationship (\S
\ref{sec:polhI}), and local interstellar clouds and the Loop I
superbubble (\S \ref{sec:loopI}).  The polarization data show that
nearby low-extinction interstellar dust, and magnetic fields, are
found in the BICEP2 region of study (\S \ref{sec:bicep}).  Conclusions
are presented in \S \ref{sec:concl}.  {Appendices provide additional
details on the stars with polarization position angles that best match
the IBEX ISMF direction that traces the LIC field (Appendix
\ref{app:hd}), the source of the color excess data (Appendix
\ref{app:ebv}), the formulae that characterize the efficiency of
polarization mechanisms (Appendix \ref{app:polhI}), and the conversion
of heliocentric velocities to the LSR for the purpose of comparing the
cloud velocity with the ISMF direction (Appendix \ref{app:lsr}). As an
aside, it is found that the contemporary solar apex motion is similar
to that found by \citet[][Appendix \ref{app:lsr}]{Herschel:1783lsr}.}

%%%%%%%%%%%%%%%%%%%%%%%%%%%%%%%%%%%%%%%%%%%%%%%%%%%%%%%%%%%%%%%%%%%%%%%%%%%%%%%%%%%%%%%%%%%%%%%%%%%%%%%
\section{Polarization data used to determine the magnetic field direction}\label{sec:data}

Starlight polarization attributed to interstellar dust aligned with
respect to the magnetic field was discovered in 1949 \citep[see][for a
review]{Andersson:2015rev}.  Multiple measurements of starlight
polarized in the interstellar medium were acquired during the last
half of the 20th century, and later assembled into a single catalog
\citep{Heiles:2000}.  The 20th century data sets used in this analysis
included the discovery of the interstellar magnetic field within 40 pc
of the Sun in the fourth galactic quadrant \citep[$\ell=270^\circ -
360^\circ$,][]{Tinbergen:1982,Piirola:1977}.  More recently, high
sensitivity polarimeters capable of $3 \sigma$ detections of
polarization strengths $<0.01\%$ have become available
\citep{Piirola:2014spie,BerdyuginPiirola:2014,PereyraMagalhaes:2007,WisniewskiMagalhaes:2007be,planetpol:2010,hippi:2015,WiktorowiczNofi:2015vesta,WiktorowiczNofi:2015alb}.

{Modern polarimeters are capable of detecting interstellar
polarizations in the low column density nearby clouds.  Surveys of
interstellar polarizations indicate that, in the absence of
line-of-sight depolarization, polarization strengths and extinction
are related \citep[][\S \ref{sec:polhI}, Appendix
\ref{app:polhI}]{Serkowski:1975ebvpol,FosalbaLazarian:2002apj}.  A
heuristic relation corresponding to the upper envelope of the
distribution of
polarization strength as a function of color-excess \ebv\ provides a nominal
upper-limit for expected polarizations. For the low
column densites of the ISM within 25 pc, \NHI$< 10^{18.7}$
\citep[][]{Woodetal:2005}, polarizations of up to $\sim 0.014$\% are
expected (see the discussion of the relation between \ebv,
polarization strengths, and $N$(H) in Appendix \ref{app:polhI}).
Modern polarimeters are therefore capable of mapping the direction of
the local ISMF.}

In this paper we utilize new polarization measurements that have been
acquired with the DiPol2 polarimeter at the KVA telescope in La Palma,
Canary Islands \citep[][]{BerdyuginPiirola:2014}, the IAGPOL
polarimeter at the LNA at Picos dos Dios in Brazil
\citep[][]{PereyraMagalhaes:2007,WisniewskiMagalhaes:2007be}, and the
POLISH2 polarimeter at the Lick Observatory in California
\citep[][]{WiktorowiczNofi:2015vesta,WiktorowiczNofi:2015alb}.  These
data include measurements with $3\sigma$ sensitivities of 0.01\% or
better.  Data on the polarizations of nearby stars from the
high-sensitivity survey with the PlanetPol instrument
\citep{planetpol:2010}, the Loop I polarization survey of
\citet{Santos:2010}, and the \citet{Heiles:2000} polarization catalog
are also incorporated into the polarization database used in this
study.

{The region of study in this paper is restricted to stars within
90\deeg\ of the heliosphere nose since this interval includes the
patch of nearby polarizing dust grains found by
\citet{Tinbergen:1982}, it includes the region of right ascension
RA$>17$ HR where \citet{planetpol:2010} have shown that polarization
strengths increase with distance, and it contains the IBEX Ribbon (\S
\ref{sec:intro}).  This region also includes the star $\alpha$ Oph (14
pc) with striking properties for the foreground ISM
\citep{MunchUnsold:1962,Frisch:1981,FrischYorkFowler:1987}.  The
angular constraint that program stars must be located within 90\deeg\
of the heliosphere nose effectively restricts stars to the galactic
center hemisphere, but includes stars at high-latitudes.  }

\begin{figure}[h!]
\plotone{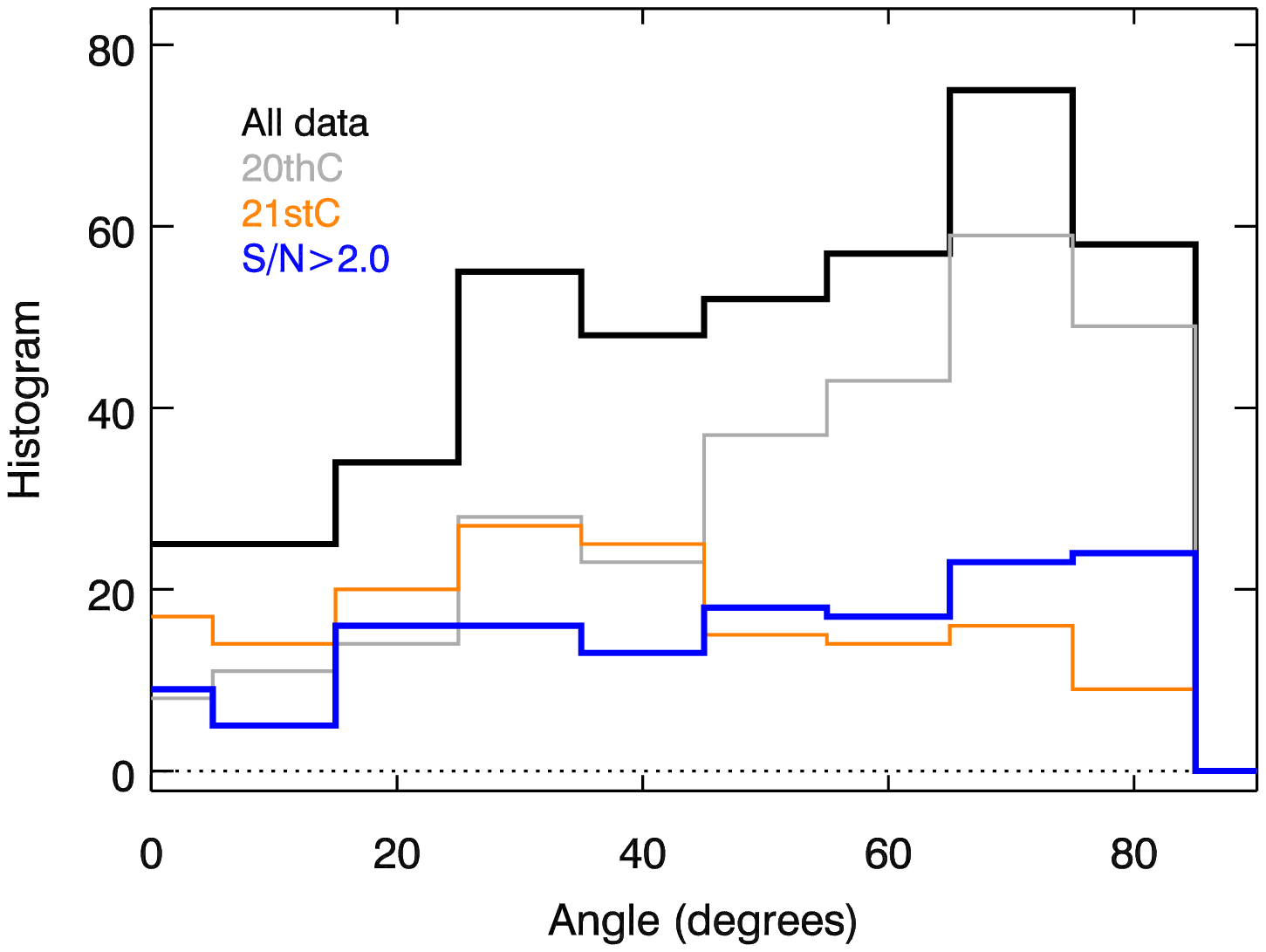}
\caption{Angular distribution of stars with respect to heliosphere
  nose direction.  The number of stars in the designated category is
  plotted against the angle between the star and the heliosphere nose.
  The thin gray and orange lines represent data collected in the 20th
  and 21st centuries, respectively.  The smaller number of 21th
  century targets at large angles from the heliosphere nose, compared
  to the same numbers for 20st century stars, is partly due to choices
  made in the selection of the target stars (Table \ref{tab:data}).
  The black line shows the total data set. The blue line shows the
  subsample with \Pol/\dPol$\ge 2.0$.  }
\label{fig:histogram}
\end{figure}

These combined data form a heterogeneous set of over 700 measurements
of polarizations for 520 stars within 40 pc.  Three hundred of these
stars are within 90\deeg\ of the direction of the heliosphere nose,
$\ell,b = 3.2^\circ,15.5^\circ$ \citep[e.g.][]{McComas:2015warmer}.
Fig. \ref{fig:histogram} histograms the distribution of the angles
between the stars in the database and the heliosphere nose as a
function of the significance of the polarization measurement and the
epoch of measurement.  Thirty percent of these stars within 40 pc have
polarizations that have been measured at a significance of
\Pol/\dPol$\ge 2.0$, where \Pol\ and \dPol\ are polarization and the
mean error of the polarization.  The merit function that we have
developed to assess the best-fitting ISMF direction to these
polarization data includes a weighting factor that allows the use of
measurements at all levels of \Pol/\dPol\ in this analysis so that
weaker polarizations are useful data points (\S \ref{sec:method}).
The data from the 20th century provide an unbiased spatial
sampling of the magnetic field (thin gray line in
Fig. \ref{fig:histogram}), whereas many of the 21st century data were
collected specifically for this project and were selected by their
proximity to the heliosphere nose.

\begin{figure}[t!]
\plotone{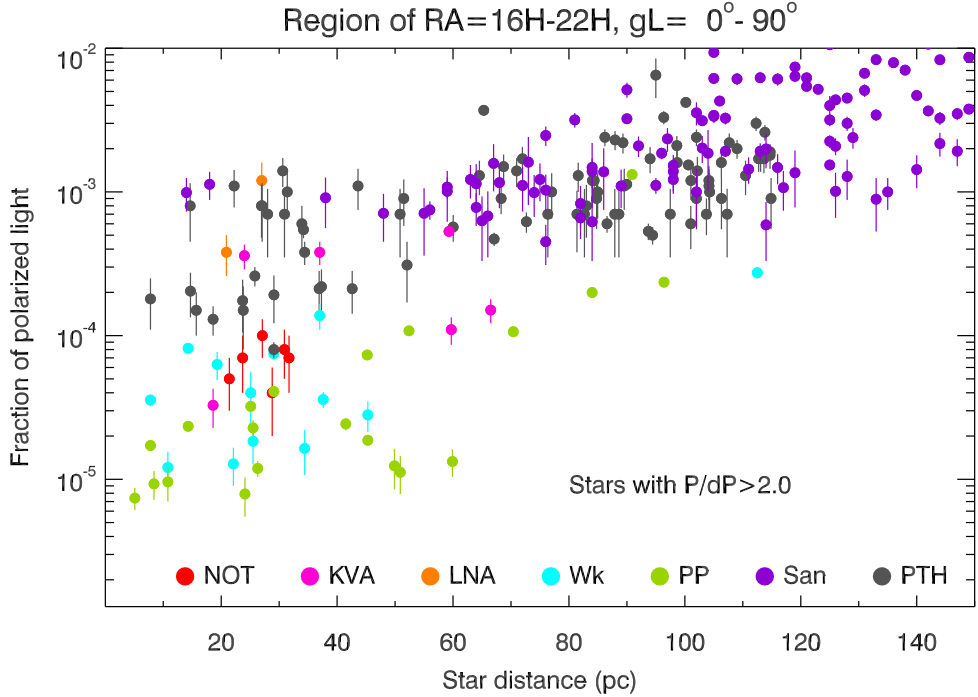}
\caption{Polarization strengths versus distance.  {The fractional
    linear polarizations of the E-component of starlight is plotted
    for stars in the first galactic quadrant, $\ell=0^\circ -
    90^\circ$, between right ascensions 17 HR and 22 HR, and where
    \Pol/\dPol$>2.0$.  The general increase of polarization strengths
    with distance becomes less obvious near the Sun where clumping of
    gas and dust becomes evident, and systematic differences due to
    instrumental sensitivities are relatively more important.  } These
  data create a heterogeneous set collected using polarimeters with
  different bandpasses and sensitivity levels.  Color coding denotes
  the data source: KVA (red), NOT (pink, Paper II), 20th century data
  \citep[PT, gray,][]{Piirola:1977,Tinbergen:1982,Heiles:2000}, LNA
  (orange), Lick \citep[turquoise,][]{WiktorowiczNofi:2015alb},
  PlanetPol \citep[green,][]{planetpol:2010}, and
  \citep[][purple]{Santos:2010}.  The distance uncertainties (not
  shown) are typically less than 4\% of the star distances.  }
\label{fig:17H}
\end{figure}

Polarizations are plotted against the star distance in
Fig. \ref{fig:17H} for those stars that are located between RA=17H and
RA=22H in the first galactic quadrant, $\ell = 0^\circ - 90^\circ$.
\citet{planetpol:2010} showed that the strengths of polarizations in
this region increase with distance (also see Fig. 7 in Paper II), and
we have shown that this increase is consistent with a dust bridge
reaching from the solar vicinity out to the North Polar Spur region
\citep{Frischetal:2014icns}.

The search for the ISMF direction that best matches these polarization
data uses polarization position angles, \PA, and not polarization
strengths (\S \ref{sec:method}).  Polarization position angles {that
trace a single magnetic field direction} are independent of the
wavelength of the measurement and therefore provide a consistently
defined quantity for data collected at different sensitivities using
different spectral bands.  In addition, the size distributions of the
polarizing grains are poorly known.  In situ measurements of
interstellar dust grains by Ulysses and other spacecraft show that the
size range extends from $\sim$0.04--2.0 \micron\ if the grains are
compact silicates
\citep{Frisch:1999,Landgrafetal:2000mass,Krueger:2014mass,Sterken:2015}.
Since the wavelength of maximum polarization strengths depend on grain
sizes, composition, and porosity
\citep{Serkowski:1975ebvpol,AnderssonPotter:2006,Andersson:2015rev},
the polarization strengths can not be compared because of the
different spectral bands of the various data sets.

\begin{figure}[t!]
\plotone{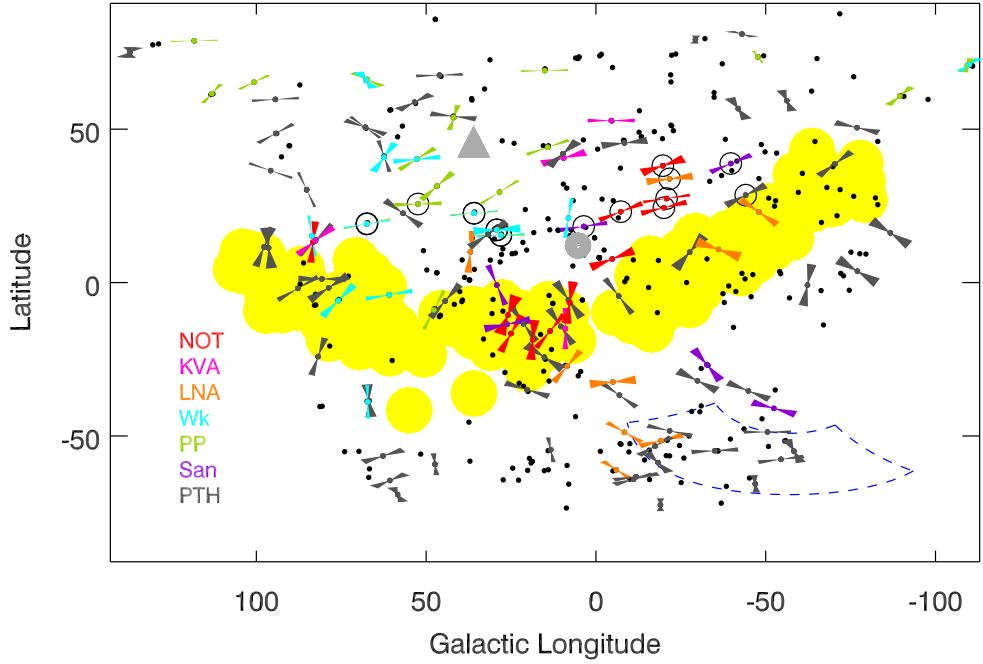}
\caption{Polarization position angles for stars with \Pol/\dPol$ >
  2.0$ and within 40 pc and 90\deeg\ of the heliosphere nose are
  plotted.  The fan-shaped polygons indicate the angular uncertainty
  of the polarization position angles \dPA.  Stars with \Pol/\dPol$<2
  \sigma$ are plotted with dots.  The regions of the highest fluxes of
  1 keV ENAs, corresponding to the IBEX Ribbon, are denoted in yellow.
  Circled stars indicate stars that trace the polarization filament
  \citep[\S \ref{sec:filament}, \S
    \ref{sec:filamentdisc}][]{Frisch:2015fil}.  The directions of the
  best-fitting ISMF \Bnofil\ and heliosphere nose are shown as gray
  triangles and dots, respectively.  The dashed lines enclose the
  region observed by BICEP for CMB B-mode polarization (\S
  \ref{sec:bicep}).  Color coding of the polarizations indicate the
  data source (see Fig. \ref{fig:17H}).  }
\label{fig:fan}
\end{figure}

The polarization position angles of the data assembled for this
analysis are mapped in Fig. \ref{fig:fan}.  The uncertainties on the
polarization position angles, arctan(\dPol/2\Pol) where \dPol\ is the
uncertainty on the polarization strength, are plotted with
``fan-shaped'' symbols.  The angular width of the fan indicates the
uncertainty on the polarization position angle.  The dots show stars
where \Pol/\dPol$<2.0$.  A nominal limit of \Pol/\dPol$=2.0$ is used
for plotting position angles since interstellar polarizations of stars
within 40 pc tend to be weak with fewer stars showing \Pol/\dPol$ >
3$.  The analysis (\S \ref{sec:method}) incorporates data with all
uncertainty levels using the appropriate weighting factor.  Since the
mean errors of the polarizations from the different data sets vary, it
is possible to have two measurements of the same star, corresponding
to two adjacent data points in Fig. \ref{fig:fan}, that show different
mean errors.  With only a few exceptions, these types of adjacent
points do not represent discrepancies but rather result from the use
of early polarization data with large uncertainties.  Generally all
measurements of a star are included with the use of weighted data
points, with the exception of a few older data points that are clearly
superceded by more accurate recent data.  The circled stars in
Fig. \ref{fig:fan} show the locations of the stars that trace a
separate nearby magnetic structure, or ``filament'' \citep[\S
\ref{sec:filament}, \S \ref{sec:filamentdisc},][]{Frisch:2015fil}.
The filament runs roughly parallel to the northern border of the IBEX
Ribbon of energetic neutral atoms
\citep[ENAs,][]{McComas:2009sci,Schwadron:2011sep}, the brightest
parts of which are plotted in yellow.

The data used in this paper differ from that of Paper II in that this
analysis utilizes new observations from the LNA and KVA observatories
(Table \ref{tab:data}), and Lick Observatory
\citep{WiktorowiczNofi:2015alb}, and excludes otherwise qualifying
Heiles data if later higher-quality data are also available.

\section{Merit function for deriving the ISMF direction from
polarization data} \label{sec:method}

Due to the low extinction of the nearby ISM where interstellar
polarization strengths are weak, we have developed a method for
combining weighted polarization data to evaluate the best-fitting ISMF
direction to those data, utilizing all data including where
\Pol/\dPol$<2.0$.  The strategy is to search for a regular dipole-like
component to the ISMF that best matches the total group of
polarization position angles indicated by the polarization data.  A
merit function describes how well the polarization position angles are
aligned with the field direction.  It is evaluated for each possible
ISMF direction to find the minimum value that provides the
best-fitting ISMF direction for the polarization dataset.

The merit function utilizes the fact that for linearly polarized
starlight aligned with the direction of the ISMF in the diffuse
interstellar medium the sine of the polarization position
angle\footnote{The polarization position angle is defined as the angle
between the linear polarization vector and a north-south meridian
passing through the star, with values in the interval 0\deeg and
180\deeg\ and increasing toward the east.}  will be zero in a
coordinate system that is aligned with the ISMF poles.  In Paper I and
II, and here, the star sample is restricted to stars within 40 pc and
90\deeg\ of the heliosphere nose.  The same strategy of minimizing the
sine of polarization position angles to evaluate the local ISMF
direction is used also, where all possible ISMF directions are tested
in order to determine the ISMF direction.  In Paper I the data sample
was restricted to \Pol/\dPol$>2.5$ (where the polarization is \Pol\
and the mean error of the measurement is \dPol).  The limit of
\Pol/\dPol$>$2.5 resulted in the omission of a large amount of useful
data.  In Paper II, using a larger set of data that included new
measurements, weighting was introduced into the merit function that
tested for the best ISMF direction, so that data points with low
statistical significance could still be incorporated into the
analysis.

This use of a weighting function is continued in the present analysis.
The weighting function is based on the bivariate statistical
description of polarization position angles given in
\citet[][NKC]{NaghizadehClarke:1993stats}.  This probability
distribution accounts for the fact that while the polarizations are
always positive, the underlying Stokes parameters can be either
positive or negative.  The outcome of using the NKC weighting function
is that the statistical likeliness of position angles in the wings of
the distribution are increased over the expectation of a true Gaussian
distribution, and this property increases the value of incorporating
polarizations with \Pol/\dPol$<2.0$ into the analysis.  Above
\Pol/\dPol=6 the probability distribution for polarization position
angles reverts to a Gaussian.

The merit function \meritfII\ that tests for the ISMF direction that best
describes the ensemble of polarization position angles, then becomes a
combination of the requirement that the mean sine of the position
angle is minimized and the statistical weighting of polarization
position angles is maximized: 
\begin{equation} \label{eqn:meritfII}
F_\mathrm{II}(B_\mathrm{i}) = \mathrm{N^{-1}}
\sum_\mathrm{n=1}^\mathrm{N} ~ f_\mathrm{n}(B_\mathrm{i})~\mathrm{where}~
f_\mathrm{n}(B_\mathrm{i})  =
\left| \frac{\mathrm{sin}(\theta_\mathrm{n}(B_\mathrm{i}))}{G_\mathrm{n}}\right| 
\end{equation} 
The contribution of each individual star \emph{n} to \meritfII,
for ISMF direction $B_\mathrm{i}$, is \meritfn.  The quantity
$\theta_\mathrm{n}(B_\mathrm{i})$ is the polarization position angle
\PA$_\mathrm{,n}$ for star n, which is calculated with respect to the
$\mathrm{i^\mathrm{th}}$ possible interstellar magnetic field
direction $B_\mathrm{i}$.  The sum is over \emph{N} stars. \Gfact\
(eqn. \ref{eqn:gfact}) is the NKC weighting factor for each star:
\noindent \newline 
\begin{equation} \label{eqn:gfact}
G_\mathrm{n}(\theta_{\rm{obs}} ;~\theta_{\rm{o}},P_{\rm{o}} ) ~ =
 \frac{1}{\sqrt{\pi}}   ~ \{ \frac{1}{\sqrt{\pi}} + \eta_{\rm{o}}  ~
\rm{exp} (\eta^2_{\rm{o}} ) ~
[1 + \rm{erf}(\eta_{\rm{o}} )] \} ~
\rm{exp}({-\frac{P^2_{\rm{o}}} {2}} )
\end{equation} 

\noindent for observed position angle $\theta_{\rm{obs}}$, ``true'' position angle
$\theta_{\rm{o}}$, $P_\mathrm{o} =\frac{P_\mathrm{true}}{\sigma}$,
mean error $\sigma$=\emph{dP},
$\eta_\mathrm{o}~=~\frac{P_\mathrm{o}}{\sqrt{2}}~\mathrm{cos}~[2(\theta_\mathrm{obs}-\theta_\mathrm{o})]$,
and the Gaussian error function $ erf (Z) ~ = ~ \frac{2}{\sqrt{\pi}}
\int_0^Z {exp}({-t^2})~dt$.

Two limits were imposed in order to prevent unrecognized properties of
the data from biasing the result.  The value of \Gfact\ was capped at
3.5 in the analysis program to prevent overweighting any single star
in the analysis. By use of this cap, the effects of possibly
unrecognized intrinsically polarized stars, or or any single data set
with systematically smaller mean errors (e.g. the PlanetPol data), are
minimized.  A second limit was imposed by requiring \Gfact\ to be $1
\times 10^{-5}$ or larger.  Experience showed that weights below this
level represent insignificant data points, but potentially cause
numerical problems in the code.  Note that polarization strengths are
not used in this analysis because of the diversity of spectral bands
with which these data were acquired (\S \ref{sec:data}).

The possible directions for the ISMF pole, $B_\mathrm{i}$, are then
tested over a grid of one-degree intervals on the sky.  For the best
comparison between the interstellar polarization data and the ISMF
traced by the IBEX Ribbon, only stars within 40 pc of the solar system
and 90\deeg\ of the heliosphere nose are included in the evaluation.
The ISMF direction determined from all qualifying data is denoted \Ball\
(see \S \ref{sec:best}).

The function \meritfII\ does not incorporate the variance of the array
being minimized.  We have therefore also tested a merit function with
the additional term, \variancemeritfII, corresponding to the standard
deviation of the array \meritfII, for the function that is being
minimized: \begin{equation}\label{eqn:meritfIII}
F_\mathrm{III}(B_\mathrm{i}) = F_\mathrm{II}(B_\mathrm{i}) \times
\sigma( F_\mathrm{II}(B_\mathrm{i})) \end{equation} We make the
assumption that if the method for determining the ISMF direction is to
yield a robust result, then the directions found from
eqs. \ref{eqn:meritfII} and \ref{eqn:meritfIII} must be similar.  This
condition is satisfied.  The use of eqn. \ref{eqn:meritfIII} does
increase the spatial gradients of the merit function, affecting the
uncertainties of the \Ball.  Results from eqn. \ref{eqn:meritfIII} are
not used to determine the best-fitting ISMF to these polarization
data.

%%%%%%%%%%%%%%%%%%%%%%%%%%%%%%%%%%%%%%%%%%%%%%%%%%%%%%%%%%%%%%%%%%%%%%%%%%%%%%%%%%%%%%%%%%%%%%%%%%%%%%%

%% \input{fig4}

\section{Direction of ISMF within 40 pc }\label{sec:result}
\subsection{Best-fitting local magnetic field direction to all data}\label{sec:best}

The simplest approximation is to assume that there is a single ISMF
that controls the optical interstellar polarizations for stars within
40 pc and 90\deeg\ of the heliosphere nose, as was also assumed in
Papers I and II.  All of the qualifying observations are used in the
evaluations of the ISMF direction, except for the 20th century data collected
in the \citet{Heiles:2000} catalog where
observations range in precision levels.  Those data are not used if a
more recent, and presumably more precise, measurement is available for the
same star.
Justification for the approximation of a single field direction is
provided by the low column densities, \NHI$< 10^{18.7}$, of the ISM
within $\sim 25$ pc \citep[e.g.][]{Woodetal:2005}, and the fact that
local clouds flow through space with roughly similar velocity vectors
in the LSR, suggesting a common origin for the clouds
\citep{FGW:2002,Frisch:2011araa}.  It is shown below, however, that
multiple magnetic structures appear to be present.

Analyzing the qualifying set of polarization data (\S \ref{sec:data}),
with the merit function \meritfII\ (eqn.~\ref{eqn:meritfII}) based on
weighted data points (eqn.~\ref{eqn:gfact}), gives a best-fitting ISMF
direction that is toward the direction \gL=16.3\deeg, \gB=27.0\deeg
(Fig. \ref{fig:meritf}, left).  The uncertainty on this direction is
determined by the width of the minimum of \meritfII.
Fig. \ref{fig:uncmeritf}, left, shows the value of \meritfII\ plotted
against the angle from this best-fitting ISMF direction for each
location in the sky.  The uncertainty on the best-fitting ISMF
direction is assumed to be the angle that clearly distinguishes the
minimum of \meritfII\ from an adjacent secondary minimum, or $\pm
15^\circ$ \cmtwo.  This new best-fitting direction differs by $27.6
\pm 29.2^\circ$ from the result of Paper II, which is not a
significant difference.  A test was also made to determine whether the
use of \meritfIII\ changed the best-fitting ISMF direction from these
data, and it gave a similar direction directed toward \gL=15.3\deeg,
\gB=27.0\deeg.

\subsection{Identifying a filamentary-shaped magnetic structure}\label{sec:filament}

As the sky coverage of the underlying polarization dataset improves,
it becomes more likely that inhomogeneities in the direction of the
local ISMF will be sampled.  Since the purpose of this study is to
connect the ISMF that shapes the heliosphere with the local
interstellar field, the analysis of the best-fitting ISMF to the
polarization data needs to take into account the possibility there are
multiple local ordered components of the magnetic field, so that
polarizations clearly associated with a magnetic field that is
different from the one that shapes the heliosphere can be omitted from
the fits.  In this section we identify a clearly identifiable
secondary magnetic structure.  The properties of this filament suggest
that it is related to interstellar dust grains deflected around the
heliosphere \citep{Frisch:2015fil}. In this section we justify the
selection of these polarizations as belonging to a separate magnetic
structure, and in the following section the magnetic field direction
in the local ISM is evaluated for a data set that omits data that
trace the polarization filament.

Using data from the PlanetPol polarimeter, \citet{planetpol:2010}
showed that the polarization strengths for stars in the region
RA$>$17H increase with the distance of the target star.  In Paper II
we showed that the stars that formed the upper envelope in the
polarization versus distance relation for this subset of the PlanetPol
data contains a group of stars within 40 pc that traces a magnetic
structure with an ordered ISMF direction that extends to within 10 pc
of the Sun.  In Paper II, this ordered field was characterized by a
position angle gradient of \PAcel\ of $\sim -0.25$ degrees per parsec
(based on the fit \PAra$= 36.0 (\pm 1.4) - 0.25 (\pm 0.03) D$ for
distance $D$ and position angles expressed in the equatorial
coordinate system, \PAra, e.g. with respect to right ascension).

Given this evidence for a magnetic structure suggested by the upper
envelope to the polarization vs. distance relation for the PlanetPol
data, we have searched for additional stars within 40 pc in this
spatial interval that might also show polarization position angles
that vary systematically with distance indicating an ordered magnetic
field.  A total of \thirteen\ stars (HD 131977, HD 161797, HD 120467,
HIP 82283, HD 119756, HD 144253, HD 130819, HD 161096, HD 134987, HD
136894) were identified in the current data set by a systematic
decrease of \PAgal\ with distance (see below).  The stars tracing the
magnetic structure are located between 6 pc and 29 pc from the Sun and
appear to form an elongated feature spanning an angle of $ \sim
5^\circ \times 98^\circ$, where the polarization position angles are
parallel to the axis of the structure (see the polarization data points that are circled
in Fig. \ref{fig:fan}). The geometric configuration of this
structure is filamentary or edge-on.  Since a filament would occupy
the smallest volume of space, we suggest that it is filamentary.

The gradient in \PAgal\ with distance is quantified by dividing the
\thirteen\ stars into two separate groups and performing a linear fit
to the variation of polarization vs. distance for each group of stars.
A slightly different slope of the \PA\ vs. distance relationship was
obtained for the two groups (Fig. \ref{fig:fit}).  The most slowly
varying ISMF component is traced by \seven\ stars and includes the
original three PlanetPol stars from Paper II.  The linear fit to
\PA\ vs. distance for the first set gives \PAgal$= 106.8~(\pm 1.5)~
-0.53~(\pm 0.08)~D_\mathrm{star}$, with a reduced $\chi^2$ of 1.217
(lower group of stars in Fig. \ref{fig:fit}).  The second set of
\five\ stars in the same extended filamentary-shaped feature can be
fit by the line \PAgal$= 130.0~(\pm 15.2)~ -0.68~(\pm
0.65)~D_\mathrm{star}$, with a reduced $\chi^2$ of 0.528 (upper group
of stars in Fig. \ref{fig:fit}).  More than \thirteen\ data points are
plotted in Fig. \ref{fig:fit} because several of these stars have been
observed multiple times.

The angle found in Paper II for the rotation of polarization position
angles with distance was $\sim - 0.25^\circ$ per parsec for position
angles presented in the equatorial coordinate system, whereas the
slopes in the galactic coordinate system are $-0.53^\circ
~\mathrm{to}~ -0.68$ degrees per parsec, and the stars span an angular
range of $\sim 98^\circ$ (\S \ref{sec:filament}).  The factor of at
least two difference between the slopes suggests that neither the
galactic coordinate system nor the equatorial coordinate system is the
correct system for evaluating the characteristics of the magnetic
structure traced by the filament stars.  It appears as if some of the
variation of the polarization position angle with distance is,
instead, due to the rotation of the coordinate systems over the
angular interval spanned by the star positions.

To remove this bias introduced by the coordinate system, we have
applied the analysis method of \S \ref{sec:method} to the filament
stars.  Evaluating the minimum of \meritfII\ for the filament-star
data gives a magnetic field direction for the filament, \Bfil, toward
\glon=359\deeg, and \glat=19\deeg.  {A better direction for the
magnetic field direction traced by the filament stars is found in
\citet[][see Table \ref{tab:summary}]{Frisch:2015fil}, which
incorporated new polarization measurements of three additional stars
not used here.}  The polarization position angles expressed with
respect to the pole \Bfil\ do not vary systematically with the
distance of the star.  Evidently the distance dependence of the
polarization position angles defined by \PAgal\ (\S
\ref{sec:filament}) is partly due to the rotation of the galactic
coordinate system over the 90\deeg\ span of the filament.

Fig. \ref{fig:fit}, right, shows the polarization position angle that
is calculated with respect to the filament magnetic field direction,
\Bfil, and plotted against the angular distance between the star and
the end of the filament. The filament end is defined by star HD
172167, located at $\ell, b = 67^\circ, 19^\circ$.  The steady
variation of the polarization angle \PAfil\ along the filament length
suggests that \Bfil\ provides a better coordinate system for
expressing filament polarizations than does the north galactic pole
(or north terrestrial pole).  The best-fitting ISMF direction to the
filament stars, from eqn. \ref{eqn:meritfII}, is directed toward the
heliosphere nose defined by the inflow velocity vector of neutral
interstellar He into the heliosphere. The magnetic turbulence
associated with the filament polarizations is $\pm 9.6^\circ$, based
on the harmonic mean of the measurement uncertainties and the
dispersion of the polarization position angles with respect to the
filament magnetic field direction.  The polarization position angles
of the filament are obviously not consistent with the local ISMF
direction obtained in Paper II, or with the new fit to the entire data
set in this paper (Fig. \ref{fig:meritf}, left).

\subsection{Best-fitting local magnetic field direction without
filament polarizations} \label{sec:domismf}

\begin{figure}[ht!]
\plottwo{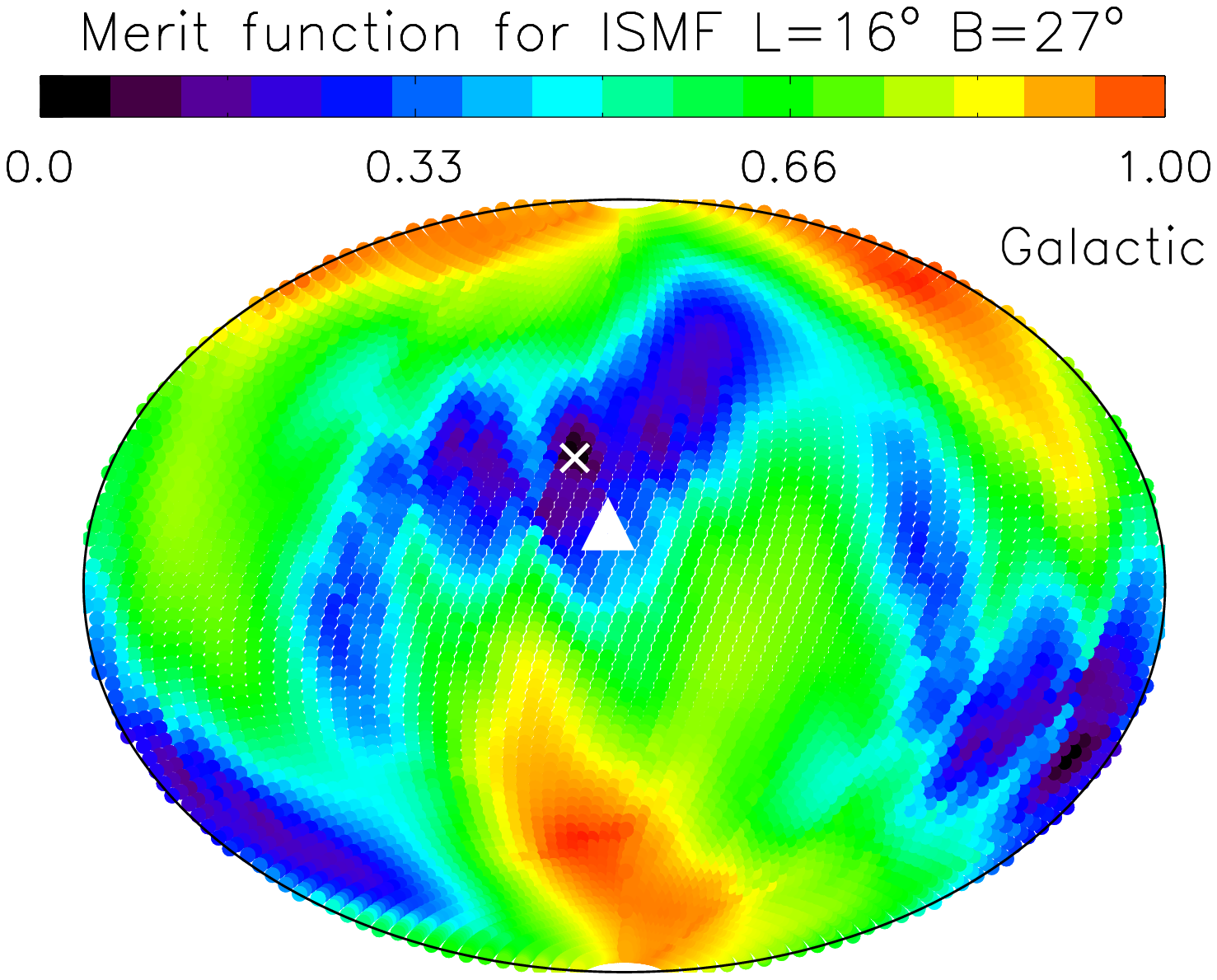}{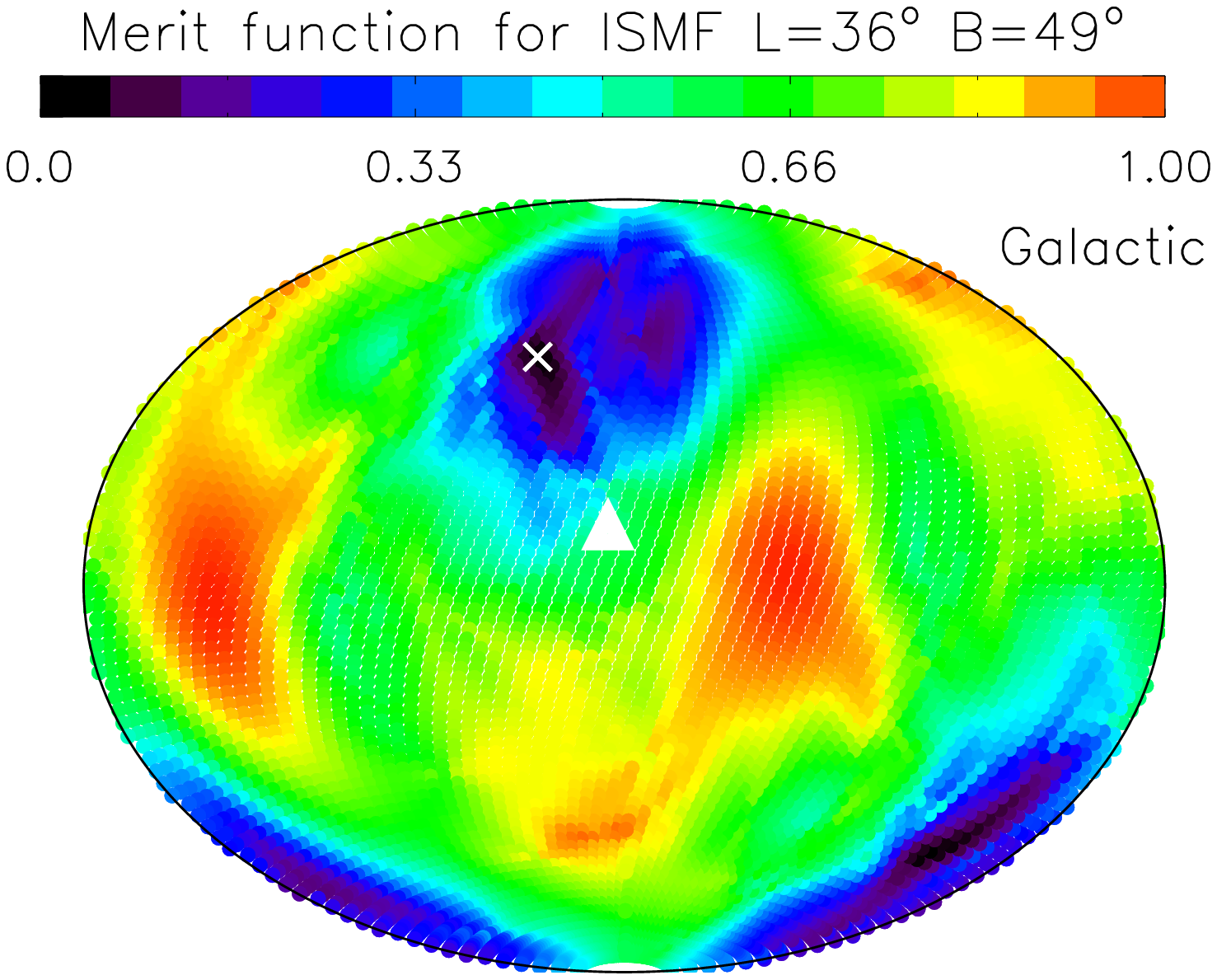}
\caption{Values of the merit function, \meritfII\ for each location on
  the sky.  \meritfII\ is normalized to the minimum value that gives
  the best-fitting ISMF direction.  The color scale is based on a log
  scale.  Left: The merit function for the fit to
  eqn. \ref{eqn:meritfII} that utilizes all qualifying polarization
  measurements.  The best-fitting ISMF direction for this star sample,
  \Ball, is toward \glon,\glat$=16.3^\circ,27.0^\circ$.  Right: The
  merit function calculated by omitting the filament stars from the
  fitted sample.  The best-fitting ISMF without the filament stars,
  \Bnofil, is toward $\ell=36.2^\circ,~b=49.0^\circ$.  The location of
  the minimum of \meritfII, is plotted with an ``X'', and the
  heliosphere nose is located at the triangle.  Uncertainties on these
  directions are shown in Fig. \ref{fig:uncmeritf} and listed in Table
  \ref{tab:summary}.  The figures are centered on the galactic center,
  with galactic longitude increasing toward the left and latitude
  increasing toward the top. }
\label{fig:meritf}
\end{figure}

\begin{figure}[t!]
\plottwo{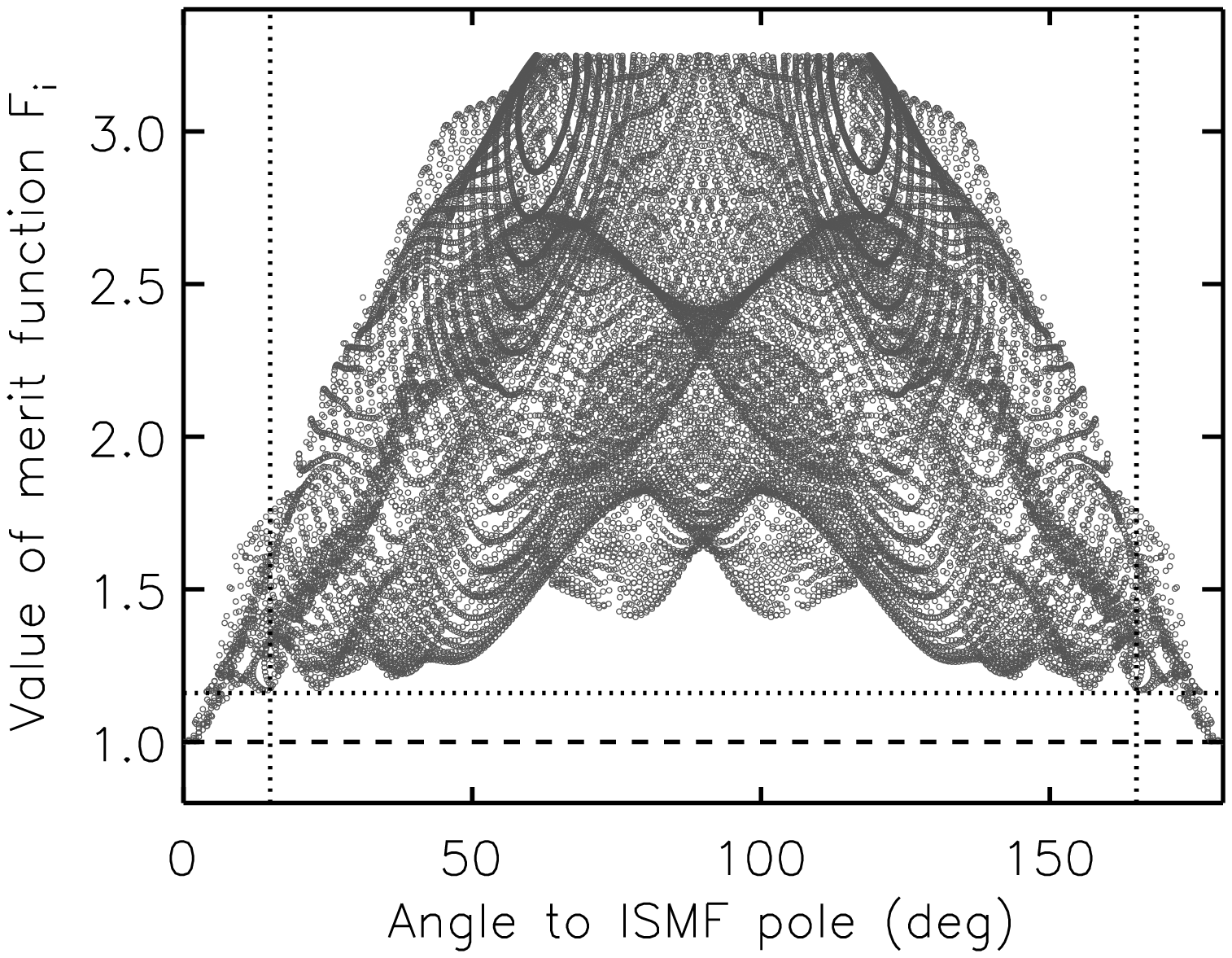}{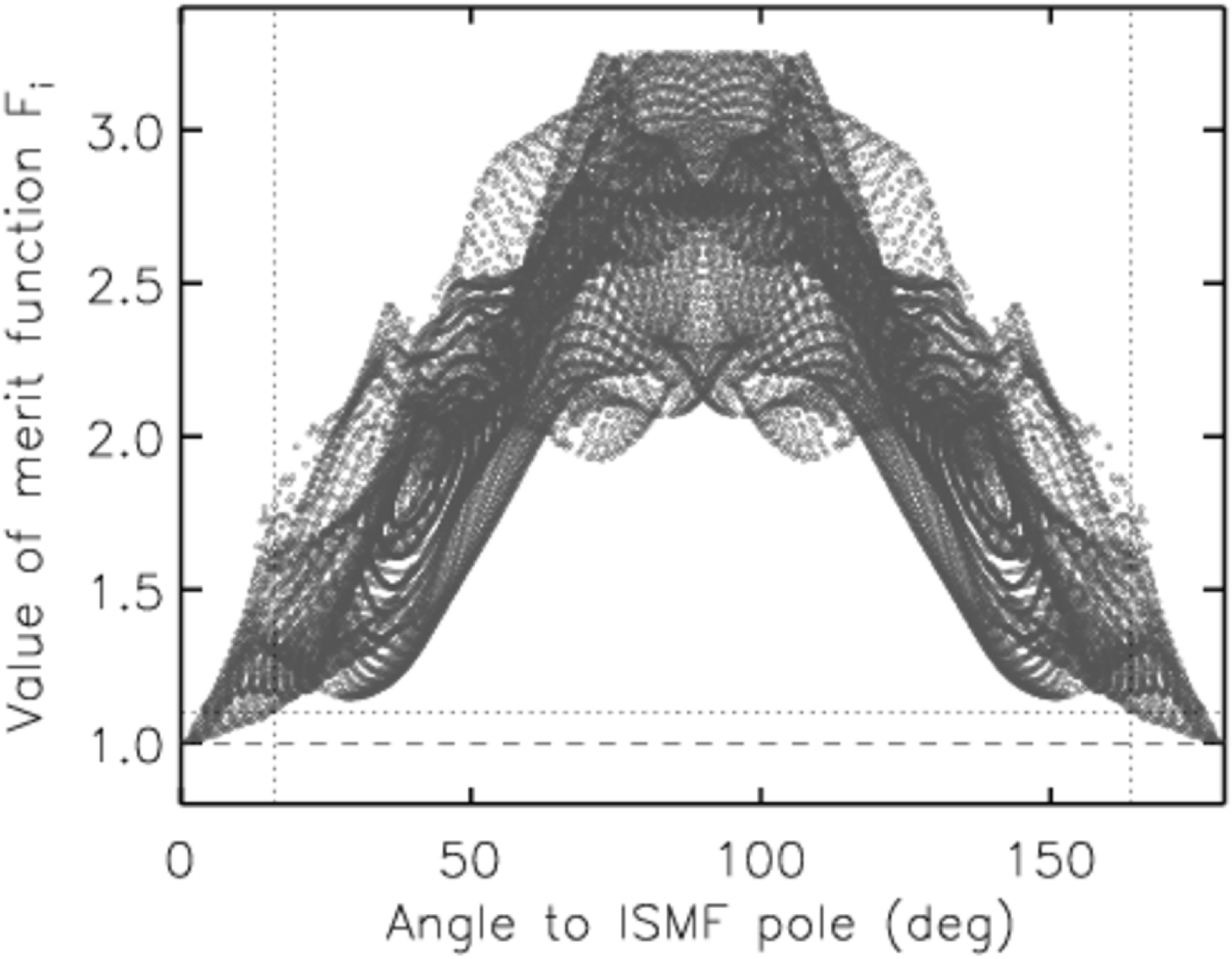}
\caption{Uncertainties on the best-fitting ISMF direction for
  \Ball\ (left) and \Bnofil\ (right).  The uncertainties are given by
  the angular distribution normalized merit function \meritfII\ values
  (vertical axis), which are plotted against the angle between the
  best-fitting ISMF direction and the merit function at each point on
  the sky (horizontal axis).  By definition, the merit function
  minimum is located at the position of the best-fitting ISMF.  For
  \Ball\ the uncertainties on the best-fitting ISMF are defined as the
  first minimum in the merit function array at $\pm 15^\circ$
  (vertical dotted line, left figure).  For \Bnofil, the uncertainty
  is arbitrarily assigned to the angle where the merit function is
  10\% above the best-fitting value, or $\pm 16^\circ$.  The values of
  \meritfn\ for the star sample with the filament stars omitted
  (right) has a more compact angular distribution and is better
  defined than the function that includes the entire polarization data
  set (left, see Fig. \ref{fig:meritf}).}
\label{fig:uncmeritf}
\end{figure}

\begin{figure}[t!]
\begin{center}
\plottwo{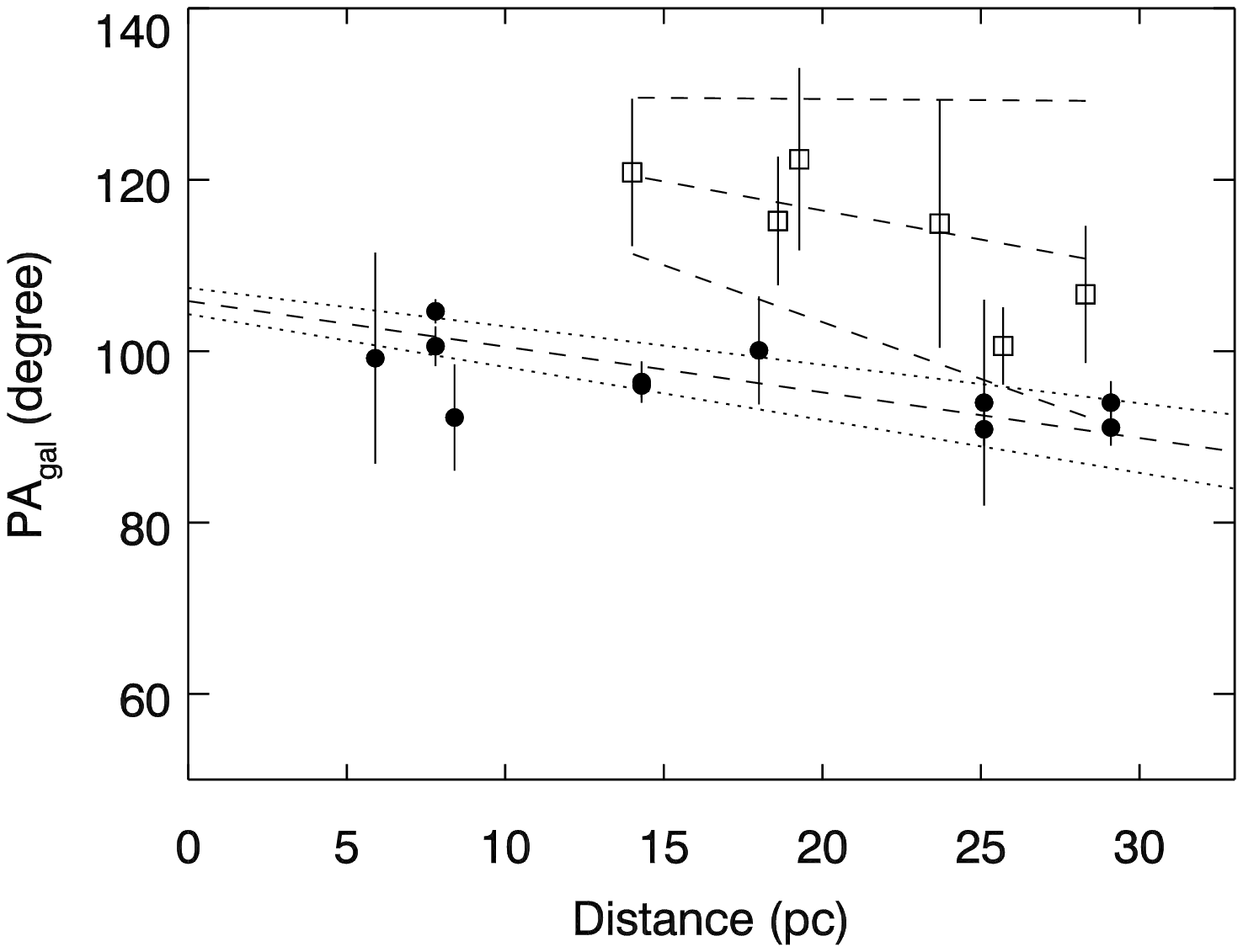}{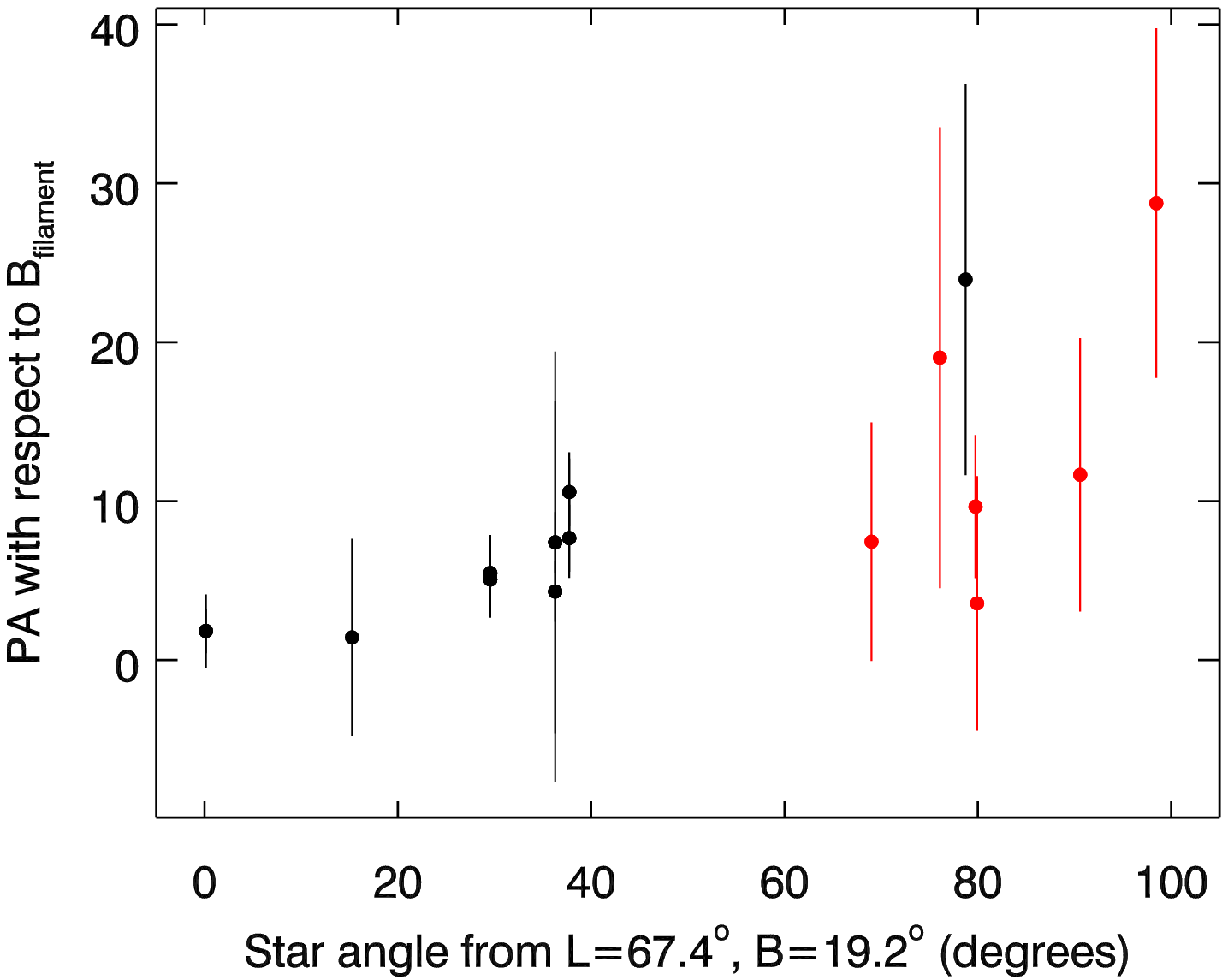}
\end{center}
\caption{Left: Galactic polarization position angles are plotted
  against the star distance for the two sets of stars that make up the
  magnetic `filament' feature (\S \ref{sec:filament}).  Both sets of
  stars show polarization position angles expressed in galactic
  coodinates, \PAgal, that rotate with the distance of the star.  The
  separate linear fits performed to the two subsets of stars in this
  filament are shown, together with the $1\sigma$ uncertainty of the
  fits.  Right: The polarization position angles (vertical axis) are
  plotted against the angular distance from the star HD 172167 that is
  located at the end of the filament (horizontal axis).  The position
  angles are expressed relative to the direction of the ISMF that
  provides the best fit to the filament polarizations, \Bfil.  The
  ISMF derived from filament polarizations provides a more uniform
  description of polarization position angles than does the north
  galactic pole.  }
\label{fig:fit}
\end{figure}

Since the filament polarizations appear to define an isolated magnetic
structure (see previous section), the fitting process has been
repeated for a data set that is identical to that used to obtain
\Ball\ except that the \thirteen\ stars that trace the filament
polarizations are omitted.  The results of the fit performed with the
omission of the filament stars are shown in Fig. \ref{fig:meritf},
right, and the uncertainties on that fit are shown in
Fig. \ref{fig:uncmeritf}, right.  For this new fit, the best-fitting
ISMF direction is toward $\ell=36.2^\circ,~b=49.0^\circ$.  Comparison
of the distribution of \meritfII\ for the evaluations with and without
the filament stars (Fig. \ref{fig:meritf}), clearly shows that the
merit function obtained from the star sample that omits the filament
stars is more clearly defined than the irregularly shaped minimum of
the merit function that is based on the entire data sample.  The
uncertainty on this best-fitting ISMF direction is assumed to occur
where \meritfII\ is 10\% larger than the minimum, giving an
uncertainty on this direction of $\pm 16^\circ$
(Fig. \ref{fig:uncmeritf}, right).  The directions of \Bnofil\ and the
IBEX ISMF, \Bibex, are the same to within the uncertainties (Table
\ref{tab:summary}).

\subsection{Statistical properties of the merit function for the
dominant ISMF, \Bnofil}\label{sec:phasespace}

At first glance, the excellent agreement between the ISMF direction
obtained from the polarization data \Bnofil\ after stars tracing a
separate magnetic structure are omitted from the sample and the ISMF
traced by the IBEX Ribbon almost seems too good to be correct.  Since
other unrecognized magnetic features may be present in these data, and
the volume of space sampled by the polarization data is large, it is
remarkable that the local ISMF field direction found from the
polarization data is so close to the ISMF indicated by the IBEX Ribbon
(Table \ref{tab:summary}).  We therefore look more closely at the
values of the individual parameters in eqn.~\ref{eqn:meritfII} to
determine whether all of the polarization data with \Pol/\dPol$>2.0$
are tracing \Bnofil, as opposed to \Bnofil\ being traced by only a
subset of the polarization data.

The properties of the polarization position angles referenced to the
best-fitting ISMF direction, \Bnofil, are viewed from two
perspectives: (i) The three-dimensional statistical properties of
\meritfII\ as a function of the probability \Gfact, sine(\PAnofil, and
\meritfII\ evaluated for the polarization position angles calculated
with respect to \Bnofil.  (ii) The probability of the position angle
(eqn. \ref{eqn:gfact}) versus the position angle of the star with
respect to \Bnofil.

\begin{figure}[t!]
\plotone{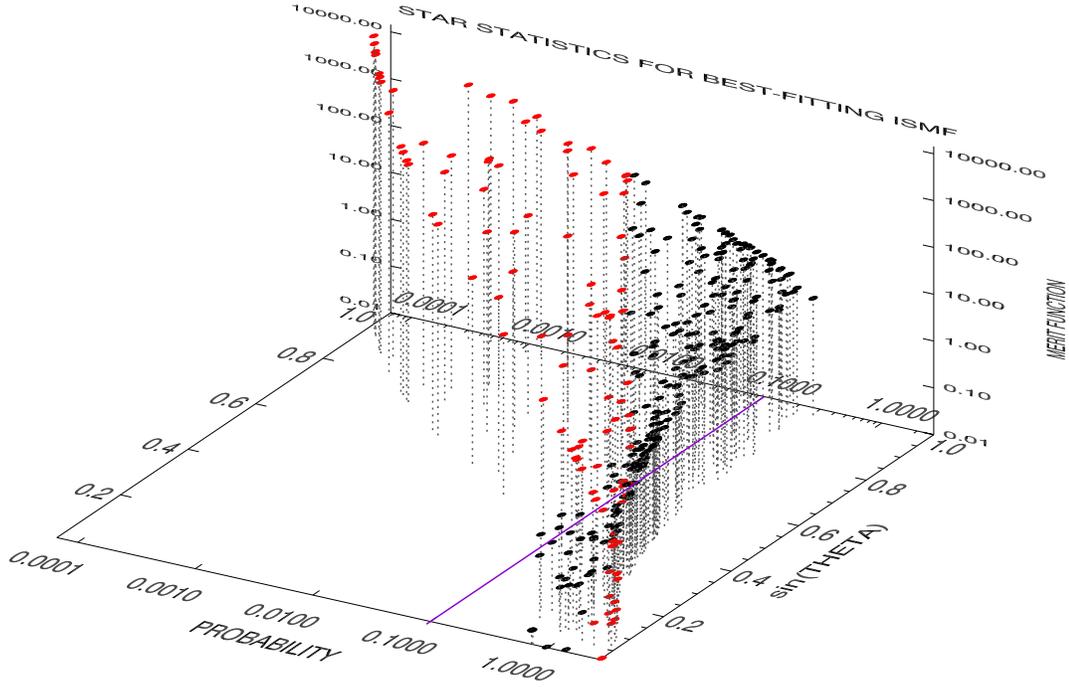}
\caption{Distribution of merit function components in 3D.  The value
  of the merit function \meritfII\ for each star determined with
  respect to \Bnofil, the best-fitting ISMF direction, is shown.  The
  front horizontal axis shows the normalized probability, \Gfact, that
  the observed polarization position angle is equal to the expected
  angle for \Bnofil\ (eqn. \ref{eqn:gfact}).  The right horizontal
  axis ``sin(theta)'' is the sine of \PA\ for each star in the
  coordinate system defined for \Bnofil\ located at the pole of the
  system.  The vertical axis, ``merit function'' (\meritfn) gives the
  merit function for each star (eqn. \ref{eqn:meritfII}).  Red points
  indicate stars with \Pol/\dPol$>2.0$.  This figure shows that the
  minimization method used to select out the best-fitting ISMF
  direction (\S \ref{sec:method}) is sensitive to stars in the
  front-right hand corner of the figure, where the values of the merit
  function being minimized are small, and those in the rear left
  corner, where the measurement errors are small but the statistical
  probability that the value corresponds to the true angle represented
  by the best-fit is small. The stars in the rear left corner are
  candidates for tracing a new component of the local ISMF structure }
\label{fig:3D}
\end{figure}

For the first approach, the statistical characteristics of the data
set that yields the best-fitting ISMF can be represented by plotting
the individual components of the merit function, \meritfII\
(eqn.~\ref{eqn:meritfII}, Fig. \ref{fig:3D}), where the best-fitting
ISMF direction \Bnofil\ corresponds to the minimum value of \meritfII.
The function \meritfn\ (eqn. \ref{eqn:meritfII}) achieves low values
for either high-probability polarization position angles (\Gfact,
eqn.~\ref{eqn:gfact}) or small \emph{sin}(\thetan).  Fig. \ref{fig:3D}
shows the \meritfII\ plotted against the probability \Gfact\ that the
data point traces \Bnofil, and the sine of the position angle
\sintheta\ calculated with respect to the best-fitting ISMF
$B_\mathrm{n}$=\Bnofil.  The probability shows the probability \Gfact\
(eqn.~\ref{eqn:gfact}) for each star \emph{i}, normalized to a maximum
value of one.  The right-hand axis labeled ``sin(theta)'' shows the
sine of the polarization position angle for each star in the
coordinate system defined with respect to the best-fitting ISMF at the
north pole.\footnote{``North'' refers to a geometric location and not
the magnetic polarity in this context.}  The vertical axis labeled
``merit function'' shows the \meritfn\ for each individual star.  Red
points show stars where \Pol/\dPol$>2.0$.  The lower statistical
probabilities of outlying polarization position angles for significant
detections (e.g. \Pol/\dPol$>2.0$) is apparent by the non-compliant
position angles in the rear-left corner of the figure.  Insignificant
polarizations would not be expected in this corner since their
position angles will tend to be statistically random so that
\sintheta$>0$.

The distribution in Fig. \ref{fig:3D} can be used to identify the set
of nearby stars with polarization position angles that are consistent
with \Bnofil. These stars include stars where measurement
uncertainties are either small, \Pol/\dPol$>2.0$, or are large but
with small values of \meritfII\ and \Pol/\dPol$<<2.0$.  Appendix
\ref{app:hd} lists the identifications of the top third of the stars
with polarization position angles that provide the best match to
\Bnofil\ and have \Pol/\dPol$>2.0$.  These stars are located on the
front right of Fig. \ref{fig:3D}.  Stars with significant polarization
position angles that do not comply with the direction of \Bnofil\ are
in the rear left part of the figure, and represent candidate
polarizations for tracing an unrecognized component of the local ISMF
(see \S \ref{sec:otherismf}).

Fig. \ref{fig:3D} shows that the minimization method used to select
out the best-fitting ISMF direction (\S \ref{sec:method}), will be
affected both by the compliant stars in the front right-hand corner of
the figure, where the values of the merit function being minimized are
small, and by the non-compliant stars in the rear left corner where
the probability that the observed polarization position angle
corresponds to the true polarization angle (given by \Bnofil) is
negligible.  The general implication of Fig. \ref{fig:3D} is that a
non-negligible fraction of the polarization position angles that are
significant (\Pol/\dPol$>2$) do not have polarization position angles
that conform to (or are compliant with) the best-fitting ISMF.  Those
points are represented by low probabilities and large \sintheta\
values in the figure.  Stars with small values of \meritfII\ are
refereed to as stars that 'conform' to, or are 'compliant' with, the
dominant ISMF direction \Bnofil.

\begin{figure}[t!]
\plotone{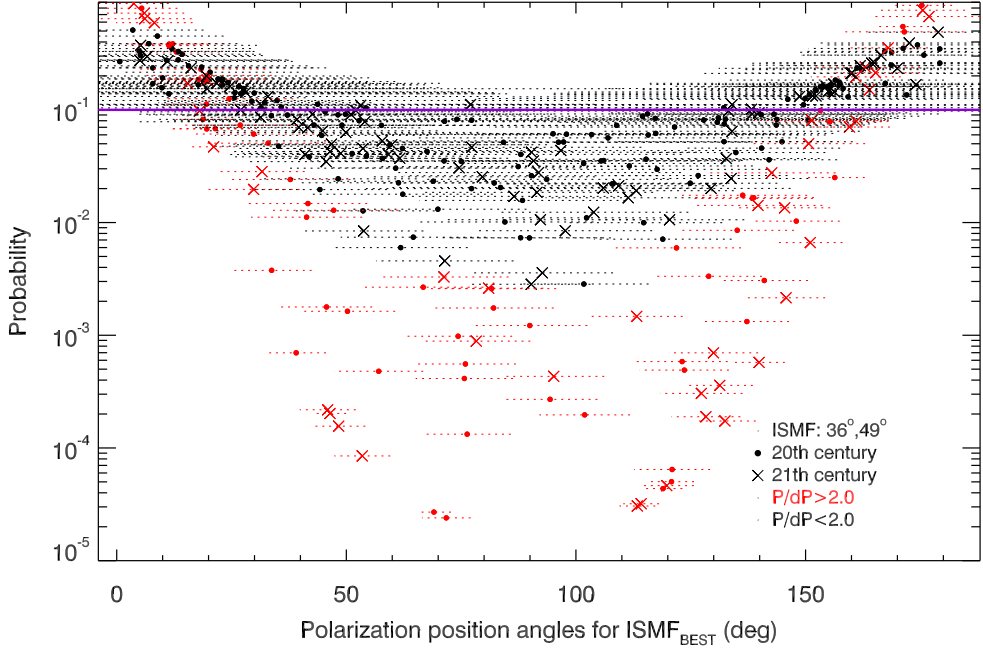}
\caption{Statistical probabilities for individual stars.  Statistical
  properties of the stars with respect to the best-fitting ISMF
  direction \Bnofil\ are plotted.  The horizontal axis shows the
  polarization position angle in the rotated coordinate frame
  corresponding to the best-fitting ISMF direction \Bnofil\ (Table
  \ref{tab:summary}).  The vertical axis gives the statistical
  probability (eqn. \ref{eqn:gfact} of the data point.  The red
  (black) points represent stars with polarization strengths of
  \Pol/\dPol\ larger (smaller) than 2.0. Data collected during the
  20th and 21th centuries are coded as ``dots'' and ``crosses'',
  respectively.  Polarization position angles that are perfectly
  aligned with \Bnofil\ have position angles of 0\deeg\ or 180\deeg.
  For reference, the purple line shows the same probability level as
  the purple line in Fig.  \ref{fig:3D}. }
\label{fig:paprob}
\end{figure}

Fig. \ref{fig:paprob} shows the probability distribution of the stars
as a function of the polarization position angle, \PAnofil, evaluated
with respect to the best-fitting ISMF, \Bnofil.  Clearly stars with
large measurement uncertainties are more likely to be compliant with
the best-fitting ISMF \Bnofil\ than stars with small measurement
uncertainties and polarization vectors pointing in the wrong
direction.  This feature allows data with all levels of accuracy to be
useful in the fitting process.  The second salient property of
Fig. \ref{fig:paprob} is that there are numerous polarization position
angles that have small mean errors and also clearly do not trace the
same magnetic field direction as \Bnofil\ and the IBEX Ribbon ISMF.

\begin{figure}[t!]
\plotone{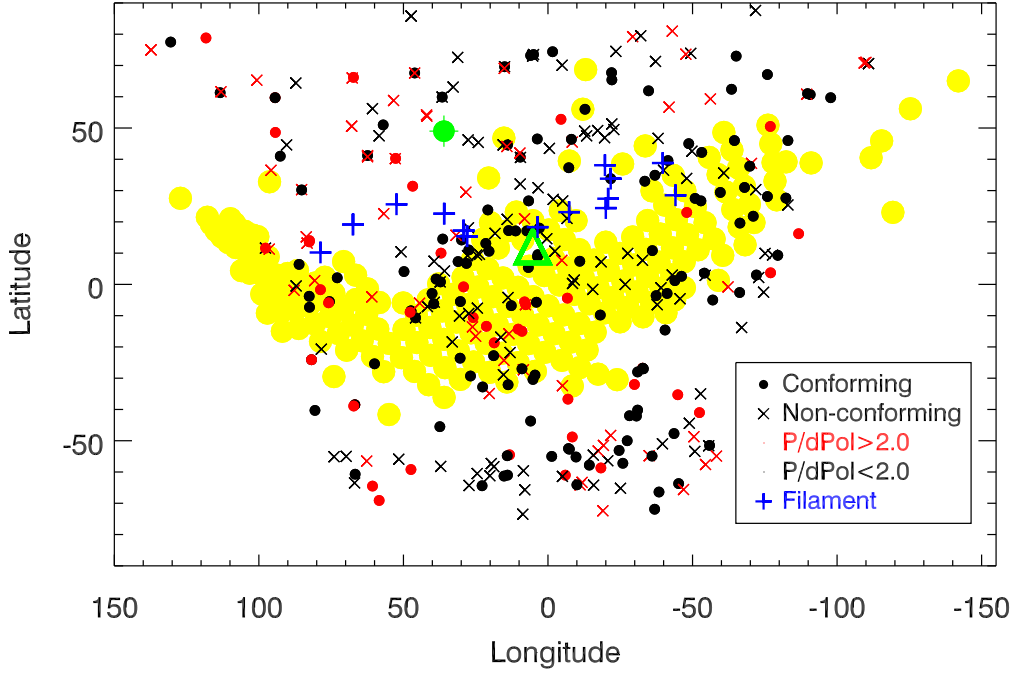}
\caption{Stars used in the study are plotted with coding that
  indicates whether or not the polarization position angles agree with
  the ISMF field direction \Bnofil.  The galactic locations of the
  stars in Fig. \ref{fig:3D} are plotted and coded by the value of
  \meritfII\ and \Pol/\dPol.  The star set is divided into two equally
  sized groups, defined by stars that are in the top or bottom half of
  stars with polarization position angles that have polarization
  position angles that agree with the direction \Bnofil.  Filled
  circles show the half of the data that best comply with \Bnofil,
  with stars having polarization mean errors of $> 2.0$ plotted in
  red, and the less significant data points plotted in black.  The
  stars with position angles that do not match \PAnofil\ are plotted
  with with ``X's''.  The distribution of position angles that do or
  do not agree with \Bnofil\ are determined using the median value
  3.54 of \meritfII.  The green triangle and green dot show the
  locations of the heliosphere nose, and the best-fitting ISMF
  \Bnofil, respectively.  The blue crosses show the locations of the
  filament stars.  Note that the stars which best trace the same ISMF
  as the IBEX ribbon tend to be more concentrated near the heliosphere
  nose and at lower galactic latitudes.}
\label{fig:2D}
\end{figure}

Fig. \ref{fig:2D} maps the stars in Fig. \ref{fig:3D} in the galactic
coordinate system, and codes the symbol of the star as to whether or
not the polarization position angle is compliant with \Bnofil.  The
star set is divided into two halves, based on the median value of
\meritfII\ of 3.54.  The best-matching half of the stars, where
\meritfII$<3.54$, are plotted with solid symbols.  The least-compliant
half of the stars, \meritfII$>3.54$ are plotted with ``X's'' (also see
Fig. \ref{fig:3D}).  Stars where \Pol/\dPol$>2.0$ are plotted with red
symbols, and stars with \Pol/\dPol$<2.0$ are plotted with black
symbols.  The compliant stars with \Pol/\dPol$>2.0$ have a tendency to
be located between galactic longitudes of 0\deeg\ and 90\deeg\ in the
northern hemisphere, and follow that trend until they wrap around
$\ell$ at negative latitudes of $\sim - 50^\circ$ in the southern
galactic hemisphere near the BICEP2 region (\S \ref{sec:bicep}).  The
conforming polarization position angles tend to follow into two
extended distributions, one located roughly between
$\ell,b=90^\circ,50^\circ$, and $60^\circ,-60^\circ$, and the other
extending roughly between $70^\circ,65^\circ$ and
$-10^\circ,-65^\circ$.  There is a slight tendency for the southern
hemisphere conforming polarization position angles to be located in
the region of the original nearby dust ''patch'' identified by
\citet{Tinbergen:1982} that extended to negative galactic latitudes in
the fourth galactic quadrant (\glon$>270$\deeg) of the galaxy.

\subsection{Turbulence of best-fitting ISMF \Bnofil }\label{sec:turb}

The turbulence in \Bnofil\ can be evaluated from the dispersion of the
polarization position angles calculated with respect to the direction
of the best-fitting ISMF direction \Bnofil.  The third of the data
sample that consists of the stars with ISMF directions that provide
the best fit to \Bnofil\ contains 114 stars.  Twenty-nine of those
stars (listed in Appendix \ref{app:hd}) have significant polarizations
with \Pol/\dPol$>2.0$.  A rough estimate of magnetic turbulence can be
found by evaluating the polarization position angles in a coordinate
system that is aligned with the pole of \Bnofil.  The average position
angle for this 29-star subset is \PAnofil$=12.0^\circ \pm 6.6^\circ$.
This relatively small dispersion about the mean suggests that
\Bnofil\ has a low level of magnetic turbulence, and is not twisted by
the kinematical properties of the CLIC (\S \ref{sec:loopI}).

A better estimate of magnetic turbulence is obtained by considering
only 21st century data that tend to have smaller mean errors than the
older data.  In the ideal case, the magnetic turbulence can be
recovered by comparing the observed position angle variations with the
mean measurement errors, or: 
\begin{equation}\label{eqn:turb}
\Phi_\mathrm{IS}^2 = \mathrm{std}(\theta_\mathrm{POL})^2
-\mathrm{std}(\delta \theta_\mathrm{me})^2 
\end{equation} 
The quantity $\Phi_\mathrm{IS}$ represents the calculated interstellar
turbulence (in degrees), $\mathrm{std}(\theta_\mathrm{POL})$ is the
standard deviation of the polarization position angle \PAnofil\
evaluated for a coordinate system with the pole located at \Bnofil,
and $\mathrm{std}(\delta \theta_\mathrm{me})$ is the standard
deviation of the mean measurement errors of the data subset.

\begin{figure}[h!t]
\begin{center}
\includegraphics[height=.3\textheight]{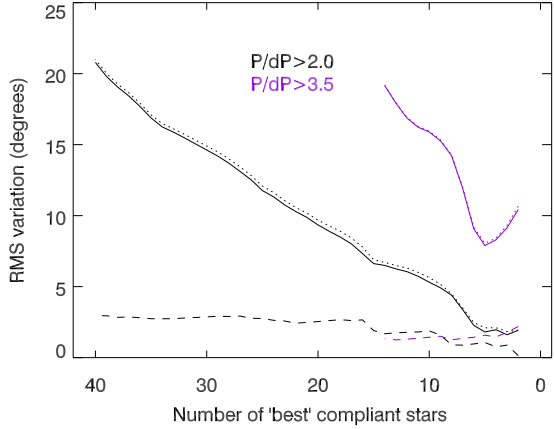}
\end{center}
\caption{Component of the position angle dispersion that can be
  attributed to interstellar turbulence.  The three quantities,
  interstellar turbulence, $\Phi_\mathrm{IS}$ (solid lines), position
  angles \PAnofil\ relative to the best-fitting ISMF direction
  \Bnofil\ (dotted lines), and mean measurement errors $\delta
  \theta_\mathrm{me}$ (dashed lines) are displayed (see
  eqn. \ref{eqn:turb}).  Interstellar turbulence is evaluated both for
  the data subset consisting of 21st century measurements and
  \Pol/dPol$>2.0$ (black), and \Pol/dPol$> 3.5$ (purple).  The
  ordering of stars along the horizontal axis is according to the
  \PAnofil, with the stars that best-agree with \Bnofil\ on the right
  of the horizontal axis, for the \Pol/\dPol\ limits above.  All
  positions on the lines represent results obtained by successively
  omitting the left-most (i.e. less compliant with \Bnofil) stars from
  the calculation of $\Phi_\mathrm{IS}$.  The minimum for the
  \Pol/dPol$> 3.5$ data brackets the best approximation for the
  interstellar magnetic turbulence indicated by these data,
  $\Phi_\mathrm{IS} \sim 9^\circ \pm 1^\circ$. }
\label{fig:turb}
\end{figure}

The amount of interstellar turbulence obtained from
eqn. \ref{eqn:turb} varies with the number of stars that are included
in the data subsample. If this subgroup of high-quality polarization
data is sorted numerically according to goodness-of-fit between
\PAnofil\ and \Bnofil, where the perfect measurement will have
\PAnofil=0\deeg, then the end-point of the array that contains the
stars with the very best matches between the polarization position
angles and \Bnofil\ should also provide the best estimate of the
interstellar magnetic turbulence using eqn. \ref{eqn:turb}.  In
Fig. \ref{fig:turb} we evaluate interstellar turbulence using
eqn. \ref{eqn:turb} and by starting with the highest quality data set
established by setting some minimum value for \Pol/\dPol. That data
subset is then reevaluated by successively rejecting the lowest
quality data points until a reasonable estimate for the interstellar
turbulence is obtained.  Fig. \ref{fig:turb} shows the estimates of
interstellar turbulence (solid lines) for two data subsets with
\Pol/\dPol$>2.0$ (black lines) and \Pol/\dPol$>3.5$ (purple lines).
The horizontal axis shows the number of qualifying stars included in
the numbers used in eqn. \ref{eqn:turb}, with the stars with
polarization position angles that better match \Bnofil\ on the figure
right, and those with poorer matches on the figure left.  For a data
subset that is restricted to stars with \Pol/\dPol$>3.5$, the minimum
value of the interstellar turbulence is 8\deeg\ and the seven stars
that bracket this minimum have $\Phi_\mathrm{IS} = 9^\circ \pm
1^\circ$.

For a larger subset where \Pol/\dPol$>2.0$, the minimum of the
interstellar turbulence decreases to $\sim 2^\circ$ for the stars that
best-comply with \Bnofil. However, the turbulent component of the
interstellar magnetic field \Bnofil\ should be best defined by the
most precise data, so we report the turbulence $\Phi_\mathrm{IS} =
9^\circ \pm 1^\circ$ from the \Pol/\dPol$>3.5$ data subset as the our
best estimate of the turbulence of \Bnofil.

\subsection{Polarization data not assigned to an ISMF structure}
\label{sec:otherismf}

The large number of non-conforming stars in Fig. \ref{fig:2D} suggests
that one or more additional ISMF directions, not yet accounted for,
must be influencing some of the polarization position angles.
\Bnofil\ is based on 343 measurements, of which 33\% have
\Pol/\dPol$>2.0$.  The set of stars with the largest third of the
values of the \meritfII\ (i.e. the one-third of the polarization data
that are least compliant with the direction \Bnofil) were selected to
be tested independently for a direction of the ISMF using the method
described in \S \ref{sec:method}.  The best-fitting ISMF direction for
this third of the stars is toward \glon=267.3\deeg, \glat=33.0\deeg,
which is 86\deeg\ from the heliosphere nose and is marginally
constrained since it is at the edge of the region that is included in
this study.

\section{Discussion}\label{sec:discussion}

\subsection{Comparing the magnetic field directions obtained from the
polarization data and the IBEX ENA Ribbon }\label{sec:ibex}

IBEX measures ENAs created from charge-exchange between neutral
interstellar atoms and heliosheath ions, including the solar wind and
incorporated pickup ions
\citep{McComas:2009sci,Livadiotis:2012innerheliosheathplasma}. IBEX
discovered an extraordinarily circular Ribbon of ENAs that is about
20\deeg\ wide and several times more intense than the distributed flux
of ENA emissions throughout the rest of the sky
\citep{McComas:2009sci,Fuselier:2009sci,Schwadron:2011sep,Funsten:2013}.
The locus of sightlines where the Ribbon is observed appear in
directions where the ISMF draping over the heliosphere is
perpendicular to the radial viewing sightline
\citep{Schwadron:2009sci}.  There is no consensus agreement on the
Ribbon formation mechanism \citep{McComasLewisSchwadron:2014}.  More
recently, \citet{SchwadronMcComas:2013retention} and
\citet{Isenberg:2014} have suggested that the Ribbon is created
through retention of pickup ions, implying that the Ribbon reflects a
true spatial structure, not an optical effect due to the prominence of
the pickup ring, as previously discussed
\citep{McComas:2009sci,Heerikhuisen:2010Ribbon}.  The Ribbon geometry
is a sensitive diagnostic of the ISMF direction and strength, and the
pressure and ionization of the interstellar cloud surrounding the
heliosphere
\citep{Frisch:2010next,HeerikhuisenPogorelov:2011,Ratkiewicz:2012Ribbon}.

The IBEX Ribbon is a highly circular feature, with a radius of
$74.5^\circ \pm 2.0^\circ$ that is centered on the ecliptic
coordinates of $\lambda = 219.2^\circ \pm 1.3^\circ, ~
\beta=39.9^\circ \pm 2.3^\circ$ \citep[based on the weighted mean
average over the energy passbands,][]{Funsten:2013}.  The Ribbon
center corresponds to galactic coordinates of \glon$=34.7^\circ \pm
4.4^\circ$, \glat$= 56.6^\circ \pm 2.6^\circ$.  The Ribbon center is
energy dependent and shifts by 9.2\deeg\ across the five energy bands
of IBEX-HI, from $\ell=34.7^\circ,~b=55.5^\circ$ for the 0.7--1.7 keV
bands, to $\ell=20.1^\circ,~b=60.7^\circ$ for the 4.3 keV band, in
galactic coordinates.

The observed center of the Ribbon arc is likely to be within 5\deeg\
of the true ISMF direction outside of the heliosphere.  MHD
simulations of the Ribbon formation by \citet{Heerikhuisen:2014newbow}
show that the center of the Ribbon arc is offset from the direction of
the ISMF far upstream of the heliosphere by 5\deeg\ for interstellar
field strengths of 3 \mG.  Comparisons between the pressures of the
inner heliosheath plasma and the thermal and ram pressure of the LIC
give an ISMF of $\sim 3.3$ \microG.  A similar value of $\sim 3.1$
\microG\ is found from the magnetic distortion of the heliotail
\citep{Schwadron:2011sep}.  Photoionization models of the LIC that
include energy sinks and sources also predict a $\sim 2.7$ \microG\
magnetic field strength from the equipartition of energy
\citep{SlavinFrisch:2008}.

The direction of the best-fitting ISMF that is obtained from the data
set that omits the filament stars, \Bnofil, agrees remarkably well
with the magnetic field direction that is obtained from the center of
the IBEX Ribbon arc.  The angular separation between the magnetic
field direction that dominates the results obtained from the
polarization data, \Bnofil, and the IBEX Ribbon field direction is
$7.6^\circ (+14.9^\circ,-7.6^\circ)$ (Table \ref{tab:summary}).  The
uncertainties become larger if the energy variation of the Ribbon
center is included.  The alignment of \Bnofil\ and \Bibex\ indicate
that these two magnetic field directions are the same to within the
uncertainties, and that the interstellar magnetic field interacting
with the heliosphere extends into the upwind interstellar regions with
minimal distortion outside of the draping region.  It is also possible
that \Bnofil\ agrees with \Bibex\ because \Bibex\ is the nearest
coherent magnetic structure in the sky and therefore has the largest
angular extent of all possible ordered fields.

Stars with polarization position angles that differ from \Bnofil, and
therefore from \Bibex, have slightly stronger polarization strengths
than those that do not agree.  However, there does not appear to be
any difference between the distances of the two subsets of the data.
The third of the polarization data that provides the worst match to
\Bnofil\ consists of 114 measurements that have a mean \Pol/\dPol\ of
$2.4 \pm 1.5$.  The third of the polarization set that provides the
best match to \Bnofil\ has a mean \Pol/\dPol\ of $1.5 \pm 1.1$.  Both
sets of stars have mean distances of approximately $25 \pm 10$ pc.
The lower mean significance of the polarizations that provide the best
match to \Bnofil, and therefore \Bibex, is not surprising since total
dust column densities will be lowest for the ISM closest to the
heliosphere.

The polarity of the ISMF is not given by either the IBEX Ribbon data
or by the polarization data.  Both the polarity of the ISMF direction
found by Voyager 1 at the heliopause \citep{Burlaga:2014ismf}, and the
radio rotation measures of pulsars within several hundred parsecs in
the fourth galactic quadrant \citep{Salvati:2010}, suggest a polarity
for the local ISMF that is directed upwards through the galactic
plane.

\subsection{Galactic cosmic ray asymmetries}\label{sec:gcr}

Asymmetries in the flux of TeV galactic cosmic rays at Earth are
observed over both large \citep{Nagashima:1998,Abdo:2009apjgcr} and
small \citep{Tibet:2011B,ARGO:2009gcr,Abbasi:2011icecubegcr} angular
scales.  The cosmic rays in the 1.5--10 TeV energy range have
gyroradii of $\sim 100-700$ AU in a 3 \microG\ magnetic field, and
probe the magnetic field in the same spatial region as the IBEX Ribbon
\citep{Schwadron:2014gcr}.  Schwadron et al. modeled galactic cosmic
ray streaming along the ISMF, for a small ratio of the
perpendicular-to-parallel component of diffusion, and showed that the
observed TeV cosmic ray asymmetries show a general ordering about the
equator of \Bibex\ locally.  Over larger spatial scales, interstellar
magnetic turbulence may disrupt GCR streaming and reduce the magnitude
of the GCR asymmetries.

The low level of magnetic turbulence found for \Bnofil,
$\Phi_\mathrm{IS} \sim 9^\circ \pm 1^\circ$, indicates that the IBEX
magnetic field extends out into interstellar space where the low
magnetic turbulence does not impede the flux of TeV GCRs into the
heliosphere.  Over spatial scales of several hundred parsecs,
\citet{Salvati:2010} used the rotation measures of radio sources to
determine a direction for the ISMF in the third galactic quadrant that
is within $\sim 22^\circ - 24^\circ$ of \Bibex.  Both the polarization
data and the radio rotation measure data suggest that the IBEX Ribbon
traces a non-turbulent magnetic field that extends into the third
galactic quadrant from whence the GCR streaming arrives.

\subsection{Possible origins of the magnetic filament} \label{sec:filamentdisc}

The origin of the filamentary structure (or structures) defined by the
polarization position angles plotted in Fig. \ref{fig:fit} is not
firmly established.  The filament polarizations trace an ISMF
direction that is aligned with the direction of the heliosphere nose,
and the filament stars are spatially arranged along a direction that
is perpendicular to the $B_\mathrm{ismf,helio} - V_\mathrm{vel,helio}$
plane that marks the heliosphere asymmetry created by the ISMF at the
heliosphere, $B_\mathrm{ismf,helio}$ and the heliocentric interstellar
gas velocity at the heliosphere, $V_\mathrm{vel,helio}$.  These
properties led to the proposal that the filament polarizations are
evidence for the deflection of the polarizing interstellar dust grains
around the heliosphere \citep{Frisch:2015fil}.  Confirmation of a
filament origin in the outer heliosheath will require modeling the
alignment and transport of interstellar dust grains during their
approach to, and interaction with, the heliosphere.  As this modeling
is not yet available, other possible origins for the filament are
briefly mentioned.

An alternate origin for the filament polarizing grains could be that
the grains are near the Sun and embedded in the LIC flow, but outside
the influence of the heliosphere.  Since the ISMF that is the best fit
to the filament polarizations coincides with the heliosphere nose
direction, which is defined by the inflowing interstellar \HeI, this
interpretation requires that either the ISMF traced by the filament is
parallel to the heliocentric LIC gas velocity, or that the assumption
of polarization vectors parallel to the magnetic field direction is
invalid.  The first requirement invokes a random coincidence between
the filament ISMF direction and the LIC heliocentric velocity and
violates the result that the LIC magnetic field and LSR velocity are
perpendicular based on IBEX data \citep[Table
\ref{tab:summary},][]{Schwadron:2015astra}.  The second requirement
may be fulfilled if radiative alignment is significant
\citep[e.g.][]{HoangLazarian:2014radiativetorques,Andersson:2015rev}.

Alternatively, if \Bfil\ is not associated with the LIC but is located
at a distance of $\sim 5$ pc from the Sun, then the filament extent is
about $1.7 \times 7.1$ pc in the plane of the sky.  If this feature
has the same density as the LIC \citep[$n \sim 0.26$ \HI\ nucleii \cc,
Model 26 in ][]{SlavinFrisch:2008}, then the column density associated
with the feature would be log \NHI=18.14 \cmtwo.  Such a column
density would be consistent with other column densities through the
very local ISM \citep{Woodetal:2005}. However, the coincidence between
the filament ISMF direction and the heliosphere nose would remain
puzzling.

The filament could be associated with an unidentified magnetic field
component, perhaps associated with a shock front related to Loop I
that extends very close to the Sun.  For this possibility, again, it
is a coincidence that the best-fitting ISMF to the filament
polarizations is toward the heliosphere nose.  The shock could be
associated with the ISM in front of one of the filament stars,
$\alpha$ Oph (HD 159561, A5 III, 14 pc), where high abundances of
refractory elements in the gas, a strong \MgI\ line indicating high
temperature or electron densities, and temperatures up to
23,000-60,000 K indicate the processing of dust through interstellar
shocks
\citep{Frisch:1981,FrischYorkFowler:1987,Frisch:1999,Crawford:2001}.
The polarizations of the star HD 159561 that traces the filament was
measured at high sensitivity by both PlanetPol \citep{planetpol:2010}
and POLISH2 \citep{WiktorowiczNofi:2015alb}, with good agreement
between the polarization position angles.

Fig. \ref{fig:ebv} compares polarization directions with the
configuration of dust reddening that is associated with the parts of
Loop I within $\sim 100$ pc (Appendix \ref{app:ebv} contains
additional information about the figure).  Some of the filament stars
have polarizations that are loosely parallel to the edge of the cavity
in the distribution of the dust extinction. {The sample of polarized
stars within 40 pc has been divided into two, and the half with
polarization position angles in best agreement with the very local
ISMF \Bfil\ are circled in green. }  The filament polarizations for
stars between $\ell=20^\circ$ and $\ell=90^\circ$ are not aligned with
strong gradients in the cumulative color excess in Fig. \ref{fig:ebv},
so a possible association between the filament and Loop I requires
further study.

\subsection{{Polarization efficiency in local ISM}} \label{sec:polhI}

{Measurements of starlight that has been polarized in the ISM provide
  one of the few viable methods for determining the distribution of
  nearby interstellar dust grains and understanding the relative
  distributions of interstellar gas and dust over parsec-sized scale
  lengths.  Polarization efficiencies in the local ISM can be found by
  comparing polarization strengths with color excess \ebv, and \ebv\
  with hydrogen column densities.  Column densities of \HI\ for stars
  within 40 pc are typically less than \NHI$\le 10^{18.7}$
  \citep{Woodetal:2005}.  Using the mean ratio between \NHI+2$N$\HH\
  and \ebv\ \citep{BohlinSavageDrake:1978}, and the upper envelope of
  the relation between polarization and \ebv\
  \citep{Serkowski:1975ebvpol} yields \ebv$\le 0.0009$ mag and
  \Polpct$\le 0.008$ for the very local ISM.  Using instead the
  analogous relations that apply to low-extinction stars with low
  column densities of foreground \HH\ predicts \ebv$\le 0.001$ mag and
  \Polpct$\le 0.014$ in the local ISM.  A quantitative discussion of
  the relations in this subsection that link hydrogen column density
  and polarization strength is given in Appendix \ref{app:polhI}.

It might be expected that interstellar polarizations of nearby stars
will be more effective than the polarizations of distant stars since
the ISM toward distant sources is likely to be more complex, with
different magnetic field directions or strong radiation fields in the
sightline depolarizing a polarized beam.  Variations in alignment
efficiency are also indicated by theoretical calculations of perfectly
aligned infinite cylinders that show approximately a factor 3--4
larger \Polpct/\ebv\ than is observed
\citep{Mathis:1979,KimMartin:1994cylin}.  The efficiency of nearby
interstellar polarization can be tested using stars where both
polarization data and hydrogen column density are available.  Two
variables are introduced to trace the efficiency of the polarization
in a sightline, $\alpha$ and $\alpha \prime$.  The ratio \Polpct/\ebv\
is proportional to $\alpha$ and $\alpha \prime$ for high-column
density and low column density sightlines, respectively.  Since \NHI\
will be used as a proxy for \ebv\ for the nearby stars (Appendix
\ref{app:polhI}) a third variable $\gamma=$\NHII/\NHI\ is included to
account for the possible presence of ionized hydrogen.

Both polarization and UV aborption line data are available for the
nearby star HD 34029 \citep[$\alpha$ Aur, Capella, 13
pc][]{Woodetal:2002capella}.  Combining UV and FUV data on Capella,
and adopting a model for the cloud length, Wood et al. determined
column densities of log \NHI$=18.24 \pm 0.07$ \cmtwo\ and log
\NHII=$18.08 \pm 0.65$ for the LIC in this direction, corresponding to
$\gamma \sim 0.69$ (with large uncertainties, $\pm 1.47 $).  The
polarization of Capella is \Pol(\%)$=0.024 \pm 0.009$
\citep{Piirola:1977}.  Combining these values with eqns. D4 and D8 in
Appendix \ref{app:polhI} for reddened and unreddened sightlines
respectively gives alpha$ \sim 5.3 $ and $\alpha \prime \sim 2.6$,
where both estimates have large uncertainties.  These large values for
polarization efficiency compared to the expected values of one for the
upper envelope of the relation between polarization and color excess
of distant stars (Appendix \ref{app:polhI}) suggest that polarization
mechanisms in the local ISM are more efficient than for distant
reddened and low-extinction stars.  These results are nevertheless
highly uncertain both because measurement uncertainties are large and
Capella is a G1III+K0III binary system where intrinsic polarization is
possible.

A second test of alignment efficiency in the local ISM is made using
the two stars with spectral types not associated with intrinsic
polarization, HD 11443 (F6IV, 20 pc) and 39587 (G0V, 9 pc) that have
polarizations of \Polpct$= 0.02 \pm 0.009$ and $0.019 \pm 0.008$
respectively \citep{Piirola:1977}.  For these stars log \NHI=18.33
\cmtwo\ and log \NHI=17.93 \cmtwo, respectively \citep{Woodetal:2005}.
Information on \HII\ toward these stars is not available so the
Capella value for $\gamma$ is assumed.  The resulting alignment
efficiencies are $\alpha=3.6$, and $\alpha \prime=1.9$ for HD 11443,
and $\alpha=8.5$ and $\alpha \prime=3.7$ for HD 39587.  If lower
ionization levels had been assumed, such as $\gamma \sim 0.29$
corresponding to the fractional ionization of the LIC near the
heliosphere \citep[Model 26 in ][]{SlavinFrisch:2008}, these values of
$\alpha$ and $ \alpha \prime$ would increase.

These estimates of the polarization efficiency in the local ISM, where
$\alpha>1$, are consistent with the results of
\citet{FosalbaLazarian:2002apj} who found a non-linear increase in
polarization as extinction approached zero for low column density
stars (see Appendix \ref{app:polhI}).  The low column density values
of $\alpha \prime>1$ found in the previous paragraph could be
spurious, resulting from the combination of the large uncertainties
and this non-linear behavior of polarization strengths at low column
densities.  Further data on both interstellar column densities and
polarizations toward the same stars are needed to establish the
efficiency of the polarization mechanisms in the local ISM.
}

\subsection{Nearby magnetic field, Loop I superbubble, and the local
interstellar cloud } \label{sec:loopI}

The agreement between \Bibex\ and \Bnofil\ (Table \ref{tab:summary})
indicates that \Bibex\ extends into the interstellar medium without
significant distortion.  The low level of magnetic turbulence for
\Bnofil\ (\S \ref{sec:turb}) suggests that the ratio of the plasma
thermal to magnetic pressure, $\beta$, is $ \le 1$.  A LIC magnetic
field strength of $\sim 3$ \microG\ is consistent with the total
interstellar pressure required to balance the inner heliosheath plasma
traced by IBEX ENAs, and the deflection of the heliotail to the port
side\footnote{The nautical terms ``starboard'' and ``port'' have been
adopted to describe the flanks of the heliosphere to the right and
left, respectively, of the heliosphere nose when referenced to the
ecliptic coordinate system. In galactic coordinates this therefore
places the starboard side of the heliosphere {\it mostly} in the
fourth galactic quadrant at positive latitudes, and the port side of
the heliosphere in the fourth galactic quadrant and {\it mostly} at
negative latitudes.} of the heliosphere \citep{Schwadron:2011sep}, and
also the equipartition of energy between the LIC thermal gas and
magnetic field \citep{SlavinFrisch:2008}. {Nearby polarized stars
with significant polarizations that do not trace \Bnofil\ (\S
\ref{sec:otherismf}) may indicate local regions where the magnetic
field has been distorted by cloud motions that create high magnetic
pressure (Fig. \ref{fig:rlpol}), or by cloud collisions
(Fig. \ref{fig:VlsrB}).}

The local magnetic field, \Bnofil, provides a probe of the physical
properties and origin of the interstellar cloudy material in the
immediate solar neighborhood.  The bulk motion through space of the
nearby interstellar material associated with the CLIC, $<30$ pc, has
been determined from the velocities for interstellar optical and
ultraviolet absorption lines and the flow of interstellar dust through
the heliosphere \citep{Frisch:2011araa}. The angle between \Bnofil\
and the upwind direction of the CLIC LSR velocity (\Vcliclsr, Appendix
\ref{app:lsr}) is $76.8^\circ (+23.5^\circ, -27.6^\circ)$.  Although
the uncertainties are large, the bulk motion of local interstellar gas
relative to the LSR is therefore perpendicular to the ISMF direction.
The flow of the CLIC relative to the LSR originates in a direction
that is within $21.4^\circ \pm 20.8^\circ$ of the nominal center of
the Loop I superbubble.  For the Loop I bubble center, we use the S1
shell feature, centered at $\ell=346^\circ \pm 5^\circ$, $b = 3^\circ
\pm 5^\circ$ as defined by \citep{Wolleben:2007}.  The S1 bubble model
places the Sun in the rim of the S1 shell.  The relative
configurations of \Bnofil, \Vcliclsr, and the S1 bubble form a
self-consistent picture where local ISM, consisting of the CLIC, is
part of the rim of the Loop I superbubble and \Bnofil\ represents the
magnetic field swept up in the rim of the expanding superbubble.

The IBEX measurements of the LIC velocity and LIC magnetic field
provide a precise set of data for comparisons between the LIC
interstellar gas velocity and magnetic field vectors.  The
heliocentric velocity determined by IBEX for interstellar \HeI\
flowing through the heliosphere corresponds to a LIC velocity with
respect to the LSR of \Vliclsr$=17.2 \pm 1.9$ \kms\ toward
\glon,\glat$=141.1^\circ \pm 5.9^\circ, 2.4^\circ \pm 4.2^\circ$
\citep[based on the velocity in ][]{Schwadron:2015isn}.  The direction
of the ISMF that shapes the heliosphere is given by the weighted mean
center of the IBEX Ribbon, at \glon$=34.8^\circ \pm 4.3^\circ$,
\glat$=56.6^\circ \pm 1.2^\circ$ \citep[Table
\ref{tab:summary},][]{Funsten:2013}.  The perpendicular angle between
the LIC velocity with respect to the LSR, and the IBEX magnetic field
direction \citep[$96.9^\circ \pm 8.5^\circ$, also
see][]{Schwadron:2014gcr} indicates that the motion of the LIC through
space is consistent with a scenario where the partially ionized LIC
\citep[Model 26 in][indicates that $\sim 22$\% of the hydrogen and
$\sim 39$\% of the helium are ionized]{SlavinFrisch:2008} sweeps up
and carries a frozen-in magnetic field through space.

Different aspects of this picture emerge when the 15-cloud model of
\citet{RLIV} is compared with the direction of the interstellar field.
The angles between the LSR velocities of the fifteen clouds
\citep[given in][]{FrischSchwadron:2014icns} and \Bibex\ are plotted
in Fig. \ref{fig:VlsrB} against \Vlsr.  The magnetic field direction
is represented by the IBEX value, as \Bibex\ is known more precisely
than the field determined from the polarization data, \Bnofil, and the
two field directions agree (Table \ref{tab:summary}).  A prominant
characteristic of Fig. \ref{fig:VlsrB} is that, except for the Aur
cloud, the LSR velocities of these clouds tend to increase as the
angle between \Vlsr\ and the interstellar field increases if the
uncertainties are included.  The mean angle between the 15 LSR
velocities and \Bibex\ is 107.5\deeg.

If we assume that deviations of the velocities of the 15 individual
clouds from the bulk CLIC motion are caused by the injection of energy
into the CLIC gas, and that the energy injection is ordered by the
pole of the IBEX ISMF direction \Bnofil\ (Table \ref{tab:summary}),
then the most effective acceleration is directed toward the antipode
of \Bnofil\ at $\ell=216^\circ,~b=-49^\circ$.  The largest deviations
from the perpendicularity of the gas and magnetic field therefore
occurs for higher velocities that are more pointed toward the third
galactic quadrant, where gas and dust densities are extremely low
\citep[e.g.][]{Fitzgerald:1968,FrischYork:1983,Vergeley:2010}.

The CLIC is a decelerating flow of interstellar gas so that cloud
collisions supply an opportunity for shock formation
\citep{FGW:2002,GryJenkins:2014,RLIV,Linskyetal:2008,Frisch:2015correcting}.
{Stars that are compliant with \Bnofil\ are preferentially located in
the first galactic quadrant, $\ell=0^\circ - 90^\circ$
(Fig. \ref{fig:rlpol}).}  Some of the stars that have polarizations
that are compliant with \Bnofil\ (\S \ref{sec:phasespace},
Fig. \ref{fig:2D}) are also located in kinematically active regions.
The elongated feature of compliant stars with $\ell = 60^\circ -
90^\circ$ (\S \ref{sec:phasespace}) occupies a kinematically quiescent
region (Fig. \ref{fig:rlpol}) dominated by the LIC, while the second
elongated feature with $\ell = -10$ to $70^\circ$ occurs in a
kinematically active region (Fig. \ref{fig:rlpol}) where multiple
clouds and large differences in absorption component velocities in the
same sightline indicate cloud collisions at up to 50 \kms\
\citep{Linskyetal:2008,RLIV}

It is notable that there is minimal nearby polarization associated
with the G-cloud centered near $\ell,b=315^\circ,0^\circ$
(Fig. \ref{fig:rlpol}, Fig. \ref{fig:gcloud}).  Two stars with
polarization position angles that comply with \Bnofil\ are found
toward the G-cloud (Fig. \ref{fig:rlpol}).  Two layers of nearby
polarizing dust are found in the direction of the G-cloud, at $\sim
19$ pc and $\sim 55-65$ pc (Fig. \ref{fig:gcloud}).  The G-cloud
extends in front of $\alpha$ Cen, at 1.3 pc, but the polarizing grains
are much more distant.

The G-cloud region shows evidence of dust grain destruction by
interstellar shocks.  The average volume densities of interstellar Fe
and Ca are systematically larger in the third and fourth galactic
quadrants where the G-cloud is located, compared to \glon$<180^\circ$
\citep{Frisch:2010s1}.  Destruction of dust grains in interstellar
shocks preferentially returns refractory elements such as Fe and Ca to
the gas phase \citep{Jones:1994,Frisch:1999}, so that the higher Fe
and Ca densities in the G-cloud region may indicate recent shock
activity and magnetic turbulence.

\subsection{Starlight polarizations in the BICEP2 field }
\label{sec:bicep}

\begin{figure}[t!]
\begin{center}
\includegraphics[height=.4\textheight]{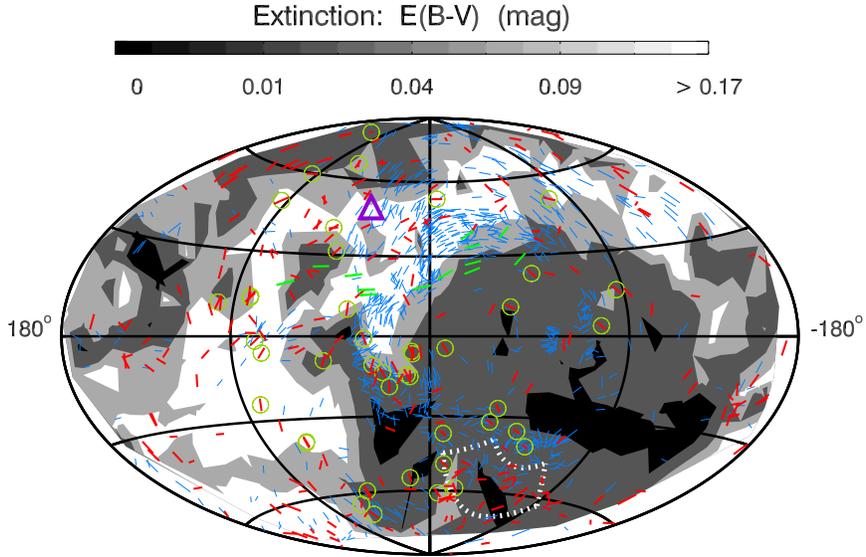}
\vspace*{-0.6in}
\end{center}
\caption{ { The amount of reddening of nearby stars is compared to
    polarized starlight. Reddening is represented by the cumulative
    smoothed color excess \ebv\ of stars within 100 pc (Appendix
    \ref{app:ebv}).  The polarizations of stars within 40 pc are
    plotted with red bars, except that the filament polarizations are
    plotted with green bars.  Nearby stars with polarization position
    angles that are in the best agreement with \Bnofil\ have green
    circles around them (see text).  The polarizations of stars within
    300 pc and that define Loop I are plotted with blue bars. The
    polarization patterns for the distant and nearby stars are similar
    in many regions suggesting that they are sampling a common
    magnetic field direction that is ordered by Loop I.  The dominant
    nearby dust structure (lightest coloring) surrounds a cavity of
    low extinction centered below the galactic plane in the fourth
    galactic quadrant (central dark region). Loop I models suggest the
    dust structure is from the expansion of Loop I into the Local
    Bubble.  The polarizations of southern hemisphere stars are more
    likely to trace \Bnofil\ than northern hemisphere stars.  The ISMF
    in the region tested by BICEP2 for the B-mode polarizations of the
    CMB (dotted lines) is different from \Bnofil.  The direction of
    \Bnofil\ is shown by the purple triangle.  }
\label{fig:ebv} }
\end{figure}

\begin{figure}[t!]
\plotone{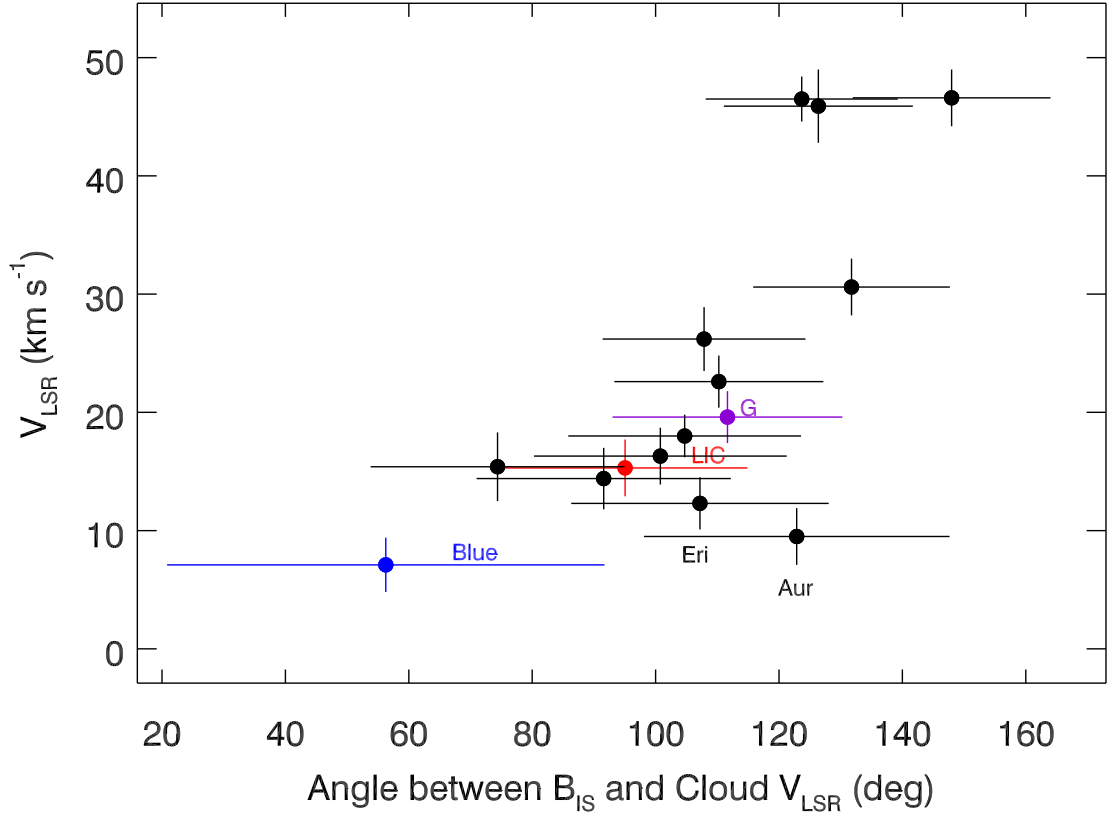}
\caption{The LSR velocities of the fifteen cloud model for the CLIC
  (see text) are plotted against the angle between the LSR velocity
  and the magnetic field direction \Bnofil.  The LSR velocity
  increases with the angle between \Vlsr\ and \Bnofil, except for the
  labeled Eri and Aur clouds.  Over half of the clouds travel at an
  angle that is quasi-perpendicular to \Bnofil\ at angles
  90\deeg-120\deeg.  Filled circles show clouds centered within
  90\deeg\ of the heliosphere nose and the LIC is plotted in red.}
\label{fig:VlsrB}
\end{figure}

\begin{figure}[t!]
\begin{center}
\includegraphics[height=.4\textheight]{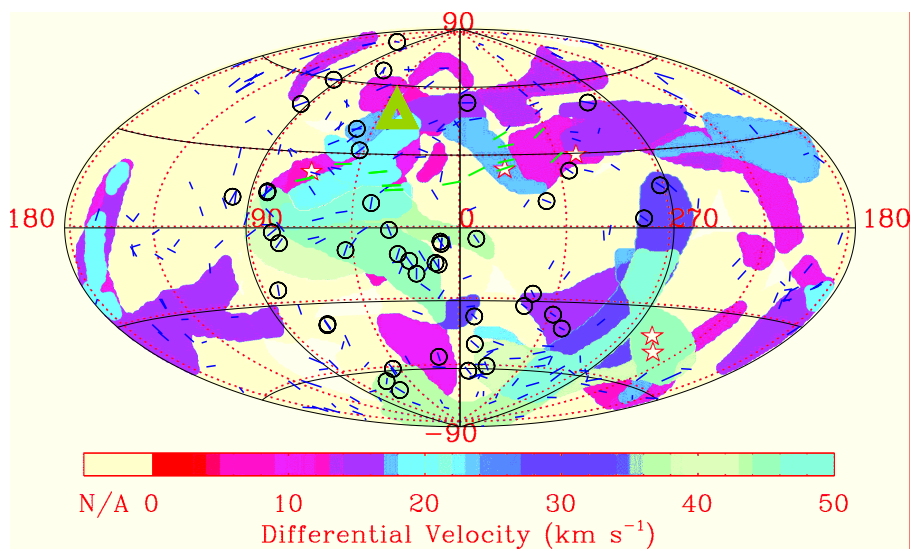}
\end{center}
\caption{{The polarizations of stars with mean errors $>2\sigma$ are
    plotted on a representation of the differential velocities between
    clouds in the same sightline.  The underlying velocity differential figure is a recolored version
    of Fig. 2 in \citet{Linskyetal:2008}.  High differential
    velocities suggest colliding clouds.}  The polarizations of stars
  that conform to \Bnofil\ are circled.  The polarization vectors are
  shown for stars with \Pol/\dPol$>2.0$, and color coded with
  non-filament/filament polarizations in blue/green.  Circled stars
  correspond to compliant stars in Fig. \ref{fig:ebv}.  \Bnofil\ is
  found in the directions of clouds with velocities that span the
  entire range of differential velocity space from 0--50 \kms, but
  favors locations in the first galactic quadrant.  }
\label{fig:rlpol}
\end{figure}

\begin{figure}[t!]
\begin{center}
\plotone{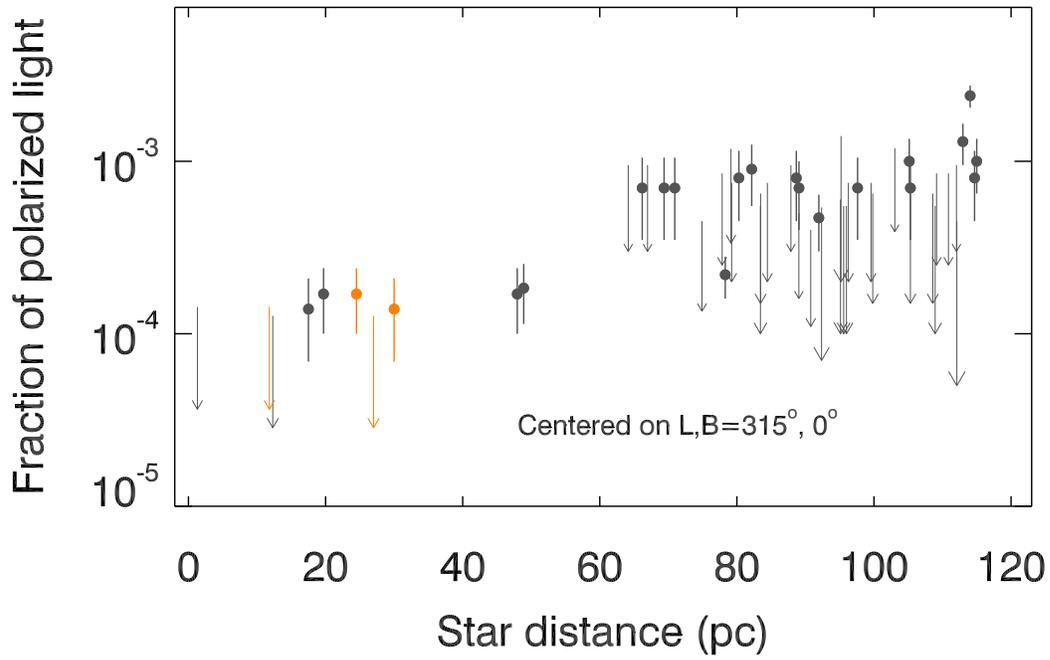}
\end{center}
\caption{ Polarizations are plotted for stars within 20\deeg\ of the
  nominal center of the G-cloud at \glon,\glat$=315^\circ,0^\circ$
  \citep[according to the model of][]{RLIV}.  Two layers of polarizing
  grains are seen, at $\sim 19$ pc and $\sim 60$ pc.  The arrows show
  upper limits on polarizations \Pol/\dPol$<2.0$.  }
\label{fig:gcloud}
\end{figure}

\begin{figure}[t!]
\begin{center}
\plotone{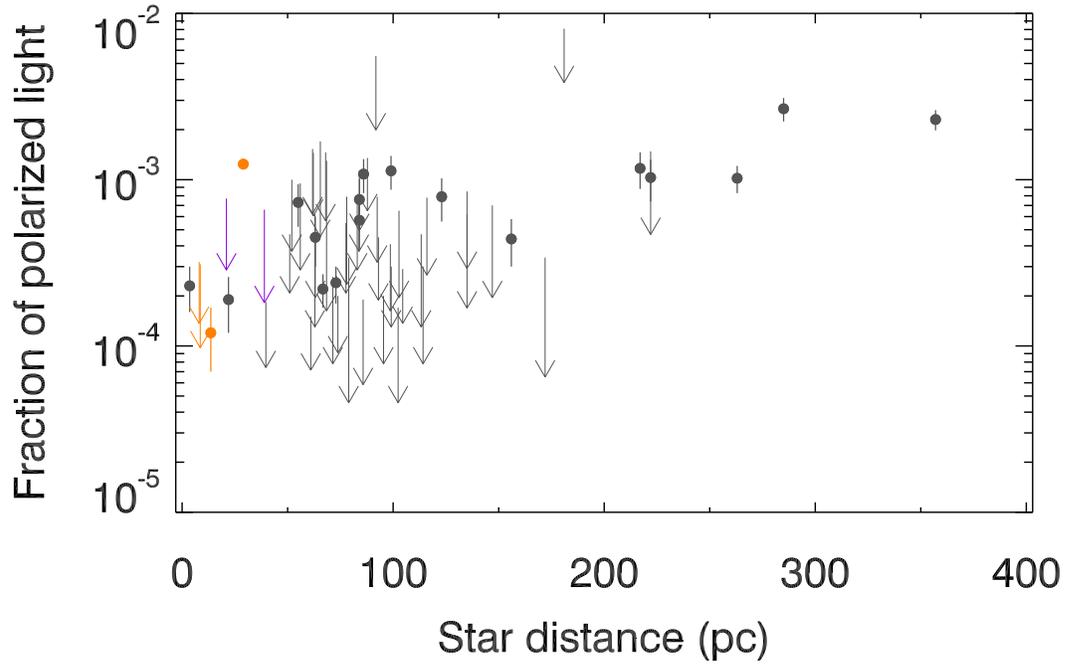}
\end{center}
\caption{Polarizations of stars located in the BICEP2 field are
  plotted against distance.  The observed polarizations suggest
  that the opacity in the B-V band in this field is $A_\mathrm{V}
  \sim 0.069$ mag or more, depending on foreground depolarization.  The
  two nearby stars with detected polarizations in the BICEP2 field and
  observed by LNA are HD 211415 and HIP 10301 (see symbol color-coding
  of Fig. \ref{fig:fan}). }
\label{fig:bicep}
\end{figure}

Interstellar polarization data provide one technique for selecting the
dust-free sightlines that provide the optimum conditions for
measurements of the B-mode polarization of the cosmic microwave
background.  BICEP2 has studied the B-mode polarization in a southern
region centered near RA=0\deeg, DEC=--57.5\deeg\ \citep{bicep2:2014}.
The BICEP2 field coincides roughly to the region defined by
declinations between --68\deeg\ and --48\deeg, and right ascensions
between 322\deeg\ and 38\deeg.  This field, centered on galactic
coordinates $\ell, b = 316.1^\circ, -58.3^\circ$, is outlined with
dotted lines on a smoothed map of color excess \ebv\ for stars within
100 pc in Fig. \ref{fig:ebv} (also see Fig. \ref{fig:fan}).
{Fig. \ref{fig:bicep} shows that the BICEP2 region contains
significant polarization within 40 pc of the Sun, and with increasing
polarization strengths out to 300 pc.  The star that is circled at the
high-longitude end of the BICEP2 region (Fig. \ref{fig:ebv}) indicates
a star with a polarization position angle that conforms to \Bnofil,
and therefore to \Bibex, which suggests that \Bibex\ extends up to the
edge of the BICEP2 region. A slightly different ISMF direction is
traced by most of polarized stars in the BICEP2 region.}

We have tested the dust content of the BICEP2 field using two markers
of interstellar dust, optically polarized starlight and the color
excess \ebv\ of stars.  The polarization data for stars in this field
are plotted against distance in Fig. \ref{fig:bicep}, using the data
sources mentioned in \S \ref{sec:data}.  Both significant polarization
detections where \Pol/\dPol$\ge2.0$, and lower polarization strengths
that are not statistically significant are found.  {Polarizations are
up to $\sim 0.2$\%, corresponding to color excess values of \ebv$\ge
0.028$ mag (using the low extinction relations of
eqn. \ref{eqn:polhIpr} for $\alpha ^\prime\le 1$), optical extinction
$A_\mathrm{V} = 3.1*$\ebv$\ge$0.087 mag for an assumed
selective-to-total extinction of 3.1, and column densities of $N$(H)$
\ge 1.4 \times 10^{20}$ \cmtwo.}  The starlight reddening data that
give the color excess values in Fig. \ref{fig:ebv} (see Appendix
\ref{app:ebv}) indicate a larger color excess of \ebv=0.19 mag through
the BICEP2 field, or optical extinction $A_\mathrm{V} = 0.59 $ mag for
a selective-to-total extinction ratio of 3.1.

Although the extinction of the BICEP2 region is negligible when
compared to molecular clouds, the polaraization data show that the
dust that is present produces detectable amounts of optical
polarization in the starlight.  Both \Bnofil\ and the lower latitudes
of Loop I extend into the BICEP2 region where they indicate that the
local contributions to polarizations are likely to trace an ordered
magnetic field with directions that can be predicted from the
measurements of the very local ISMF \Bnofil\ and the southerly portion
of Loop I.

\section{Conclusions}\label{sec:concl}

Starlight that is linearly polarized while traversing the dichroic
local ISM is utilized to chart the direction of the magnetic field
within 40 pc and 90\deeg\ of the nose of the heliosphere,
corresponding roughly to the galactic center hemisphere.  The purpose
of the study is to compare the direction of the ISMF that shapes the
heliosphere, and is traced by the unconventional magnetic field
diagnostic provided by the IBEX ENA Ribbon, with the local ISMF
direction traced by the polarization data.

New data on polarized starlight acquired in the northern and southern
hemispheres provide the basis for this study.  These polarization
measurements, with typical $3\sigma$ sensitivities of 0.01\%, have
been collected with telescopes at the KVA, LNA, and Lick
observatories.  Using the new polarization data and additional data
from Paper II and the literature, we determine several nearby magnetic
structures, one of which coincides in direction with the interstellar
magnetic field traced by the IBEX Ribbon.  Summarizing the results:

\begin{itemize}

\item A merit function, \meritfII, has been developed for evaluating
the magnetic field direction that best fits the polarization data.
\meritfII\ assumes that the linear interstellar polarizations are
parallel to the magnetic field direction.  Evaluation of the merit
function for the entire qualifying set of polarization data, within 40
pc and 90\deeg\ of the heliosphere nose direction, results in a local
ISMF direction in agreement with earlier values in Paper II.

\item A visually and numerically distinct magnetic filament is traced
by the polarizations of \thirteen\ stars, of which the nearest star is
within six parsecs.  These filament stars were originally selected
based on a gradient of the polarization position angles with distance,
for position angles given in the galactic coordinate system, but that
effect was found to be spurious (\S \ref{sec:filament}).  Utilizing
the function \meritfII\ to obtain the ISMF direction that is traced by
the filament polarizations gives an ISMF direction that is located
within $15^\circ \pm 10.3^\circ$ of the heliosphere
nose. \citet{Frisch:2015fil} favor an origin for the filament
polarizations related to the interstellar dust that is deflected
around the heliosphere in the outer heliosheath. Other possibilities
are that the feature is associated with the local ISM or possibly
related to the near side of Loop I.

\item The ISMF that shapes the heliosphere, \Bibex, has been
identified in linearly polarized starlight for the first time.  When
the filament stars are omitted from the polarization data set that is evaluated
with the merit function \meritfII, the ISMF direction of \Bnofil\ is aligned with
the direction $\ell=36.2^\circ  , b=49.0^\circ  ~ (\pm 16^\circ)$.
The direction of \Bnofil\ is centered 7.6\deeg\ (+14.9\deeg, --7.6\deeg) away
from the ISMF direction found from the IBEX Ribbon (Table
\ref{tab:summary}).  The polarization position angles of the stars
that best comply with \Bnofil\ indicate that the magnetic turbulence
of \Bnofil\ is weak (\S \ref{sec:turb}).  Several of the stars that
trace \Bnofil\ are located within $\sim 10$ pc.  \Bnofil\ is not
distinguished by the mean distance of the stars that trace \Bnofil,
but those stars tend to show weaker polarizations than average.

\item The local ISMF must be complex because many stars have
polarization position angles that trace a magnetic field direction
that is different from \Bnofil.

\item \Bnofil\ must thread the local interstellar clouds.  It forms an
angle of $76.8^\circ$ $(+23.5^\circ, -27.6^\circ)$ with the bulk CLIC
motion relative to the LSR.  This result agrees with previous findings
and is consistent with a model where the CLIC is associated with an
evolved rim of the Loop I superbubble.

\item The velocities of the 15 CLIC clouds identified by \citet{RLIV}
have been converted into the LSR; those LSR velocities increase as the
angle between the LSR velocity and \Bnofil\ increases.  This suggests
that \Bnofil\ orders the kinematics of the local ISM.  One possible
scenario is that the polarized dust bridge extending from the
heliosphere to the North Polar Spur region represents dust and the
magnetic fields swept up by the large-scale expansion of the Loop I
superbubble.

\item { The polarizations of three stars within 20 pc have been
compared with hydrogen column densities.  Values of \ebv\ were
estimated from \NHI\ and then compared with polarization strengths (\S
\ref{sec:polhI}). The resulting ratios \Polpct/\ebv\ are consisent
with the results of \citet{FosalbaLazarian:2002apj} that find an
upturn in polarization strengths for low-extinction stars, although
uncertainties are large.  }

\item The polarization data indicate that \Bnofil, which coincides
with \Bibex, extends to the edge of the region that was tested by
BICEP2 for the B-mode polarization of the CMB.  Polarizations of up to
0.2\% inside of the BICEP2 field are found, corresponding to a color
excess of \ebv$\ge$0.028 mag.  Data on starlight reddening give a
larger color excess, \ebv=0.19 mag in the BICEP2 field, corresponding
to an optical extinction $A_\mathrm{V} = 0.59 $ mag.  \end{itemize}

{The only nearby interstellar cloud where the relation between the gas
and dust is clearly established is the LIC, where multiple spacecraft
have measured UV LIC absorption lines and the ISMF direction has been
found from heliosphere models and the IBEX Ribbon.  Achieving a
similar understanding of the relation between the local ISMF traced by
the polarization data and interstellar clouds will require UV studies
of the interstellar absorption lines toward the same stars for which
high-quality polarization data are available.  Only then will a full
understanding of the relation between
the ISMF that shapes the heliosphere and the
magnetic field that is associated with interstellar clouds in the solar
vicinity be possible.}

\acknowledgements This research has been partly supported by the NASA
Explorer program through support for the IBEX mission, and by the
European Research Council Advanced Grant HotMol (ERC-2011-AdG 291659).
P. Frisch would like to thank Stephen Case for pointing out the first
study of the solar apex motion in \citet{Herschel:1783lsr}.  \appendix

%% Appendix A 

\section{Appendix: Stars with polarizations that best conform to
\Bnofil } \label{app:hd}

The 29 stars with \Pol/\dPol$\ge2.0$ that belong to the third of the
sample with polarization position angles that best match \Bnofil\ (see
\S \ref{sec:phasespace}) are: HD 11276, HD 90132, HD 90355, HD 91324,
HD 112413, HD 117939, HD 126660, HD 127762, HD 130109, HD 150680, HD
161892, HD 169916, HD 173818, HD 177409, HD 177716, HD 184509, HD
185395, HD 187642, HD 190248, HD 197989, HD 198149, HD 205478, HD
207129, HD 210027, HD 210049, HD 210418, HD 215696, HD 216435, HD
223889.

%%  Appendix B

\section{Appendix: Color excess \ebv } \label{app:ebv}

The color excess \ebv\ contours in Fig. \ref{fig:ebv} are based on the
photometric and astrometric data for stars brighter than V=9 mag in
the Hipparcos catalog \citep{Perrymanetal:1997}.  Color excess values
\ebv\ are calculated using the intrinsic stellar colors as a function
of spectral type given by \citet{Cox:2000}.  Poorly defined spectral
types are weeded out by only using stars where astrometric distances
match photometric distances to within 15\%.  The uncertainties on the
astrometric distances are used to spatially smooth the \ebv\ values
over $\pm 13^\circ$ angles in the sky for stars with astrometric
distances that overlap.  Variable stars are not included in the
construction of the \ebv\ maps.  Variability is filtered out by
excluding Hipparcos data with variabilities that are larger than 0.06
mag.  {The contour levels of \ebv = 0.01, 0.04, 0.09, and 0.17 mag
correspond to foreground hydrogen column densities of 19.76, 20.37,
20.72, and 20.99 \cmtwo\ for N(H$^o$+2H$_2$)/\ebv$=5.8 \times 10^{21}$
\cmtwo\ mag$^{-1}$ \citep{BohlinSavageDrake:1978} providing that
negligible amounts of \HII\ are present.}

%%  Appendix C

\section{Appendix: Column densities, color excess and polarization
strengths} \label{app:polhI} 

{This appendix summarizes the relations between hydrogen column
densities and polarization strengths as determined from the
literature.  The discussion in \S \ref{sec:polhI} uses these relations
to evaluate whether the polarizing mechanisms in the local ISM are
more efficient than in the generic ISM.

The relations between color excess \ebv\ and polarization strengths
has been determined both for reddened stars
\citep{Serkowski:1975ebvpol} and lightly reddened stars
\citep{FosalbaLazarian:2002apj}.  The upper envelope of the plot of
color excess \ebv\ vs.polarization strength for generic reddened
stars within several kpc of the Sun is given by \Polpct=9 \ebv, where
\Polpct\ is percent polarization \citep[][\S
\ref{sec:data}]{Serkowski:1975ebvpol}.  The upper envelope of the
relation between polarization and color excess is found to be
non-linear for lightly reddened stars, \ebv$< 1$ mag, giving the
alternate relation \Polpctpr=3.5 \ebv $^{0.8}$ \citep[eqn. 3 in
][]{FosalbaLazarian:2002apj}.  The term ``upper envelope'' indicates
the maximum polarization that has been observed as a function of
extinction, and therefore by common assumption the maximum allowable
polarization for a given level of extinction.

Obtaining the dependence of polarization on hydrogen column densities
requires the use of an additional relation between extinction and
column density.  The \emph{Copernicus} mission studied the absorption
lines of nearby stars in the far-UV where the \HI\ Lyman lines and
\HH\ absorption lines are located
\citep{SavageBohlin:1977H2h2,BohlinSavageDrake:1978}.  Observations of
\HI\ and \HH\ yielded a mean ratio of total neutral hydrogen to color
excess of $< N ( \mathrm{H}^\circ + \mathrm{H_2})>/$\ebv$= 5.8 \times
10^{21}$ \cmtwo\ mag$^{-1}$.  \emph{Copernicus} did not directly
measure column densities of ionized hydrogen, \NHII.  The ionization
corrections needed to obtain total hydrogen column densities obtained
from $\mathrm{H}^\circ + \mathrm{H_2}$ are expected to be less then a
few percent for the \emph{Copernicus} stars
\citep{BohlinSavageDrake:1978}, nearly all of which are beyond 100 pc
and therefore beyond the boundaries of the Local Bubble that is nearly
devoid of interstellar dust \citep{Fitzgerald:1968}.

Observed polarizations may differ from the relation expected from the
upper envelope of the relation between polarization and extinction. We
introduce the terms $\alpha$ and $\alpha\prime$ to evaluate the
polarization efficiency of reddened and unreddened sightlines,
respectively.  { Variations in the terms $\alpha$ and $\alpha\prime$
can result from the presence of several different magnetic field
directions foreground to the star, patchy dust distributions, and/or
variations in the grain characteristics.  }

The total hydrogen column density is used as a proxy for extinction
because the extinction toward nearby stars is too low to be determined
with the standard evaluation of the difference between the attenuation
of light in the B and V passbands that creates the color excess \ebv.
Hence a second factor that will influence the evaluation of the
efficiency of alignment mechanisms is the fractional ionization of
hydrogen.  Ionization corrections are implemented through the term
$\gamma = N(\mathrm{H}^+) /N(\mathrm{H}^\circ) $, where \NHII\ is the
column density of ionized hydrogen.

Significant amounts of \HH\ are not expected in the local ISM because
of high fluxes of far-UV and extreme-UV ionizing radiation
\citep{Vallerga:1998,VallergaSlavin:1998} and low column
densities\citep{Woodetal:2005}.  Locally the high EUV fluxes also
generate significant amounts of \HII\ in the LIC
\citep{SlavinFrisch:2008} and other clouds
\citep{SlavinFrisch:1998,RedfieldFalcon:2008}.  The term
\NHI+2\NHH\ in \citep{BohlinSavageDrake:1978} can therefore be
replaced by \NHI+\NHII=\NHI(1+$\gamma$) for the very local ISM.

For reddened stars and the mean $N$(H)/\ebv\ \emph{Copernicus} relation:
\begin{eqnarray}\label{eqn:polhI}
 \mathrm{P_\%} = \alpha ~ 9 ~ \mathrm{E(B-V)} \\ 
 C = 1.55 \times 10^{-21} ~ \mathrm{mag~cm^2} \\
 \mathrm{P_\%} = \alpha ~ C ~ (N\mathrm{(H^\circ)}  + N\mathrm{(H^+)}) \\ 
 \alpha = \frac{  \mathrm{P_\%}  }{ C ~ N\mathrm{(H^\circ)} ~ (1+\gamma)  } 
\end{eqnarray}

Low extinction stars have a slightly different ratio of hydrogen
column density to \ebv.  Restricting the \emph{Copernicus} data sample
to those stars with a small fraction of \HH, $<1\%$, provided a sample
of the ``intercloud medium'' where the mean ratio between H and
\ebv\ is $< N ( \mathrm{H}^\circ + \mathrm{H_2})>/$\ebv$= 5.0 \times
10^{21}$ \cmtwo\ mag$^{-1}$ \citep{BohlinSavageDrake:1978}.  For
lightly reddened stars and the \emph{Copernicus} relation for the
intercloud medium the polarization-extinction measure becomes:
\begin{eqnarray}\label{eqn:polhIpr}
 \mathrm{P_\% ^\prime} = \alpha^\prime ~ 3.5 ~ \mathrm{E(B-V)}^{0.8} \\ 
 C^\prime = 1.53 \times 10^{-17} ~ \mathrm{mag~cm^2} \\
 \mathrm{P_\%} = \alpha^\prime ~ C^\prime ~ (N\mathrm{(H^\circ)}  + N\mathrm{(H^+)^{0.8}}) \\ 
\alpha^\prime = \frac{\mathrm{P_\%} }{ C^\prime ~ N\mathrm{(H^\circ)}^{0.8}  ~ (1 + \gamma)^{0.8}} 
\end{eqnarray}
which can be used to estimate both $\alpha^\prime$ and $\gamma$ for
sightlines where \HI\ and \HII\ and polarization data are available,
where $\alpha \prime$ is the efficiency of grain alignment in the
absence of depolarization effects.

These relations are used to evaluate the polarization efficiency in
the local ISM in \S \ref{sec:polhI}.

}  %%  end pcfbold

%%  Appendix D

\section{Appendix: Bulk motion of local interstellar clouds relative
to the LSR } \label{app:lsr}

{Astrometric data collected by the founders of modern astronomy
revealed that both the Sun and stars are moving through space
\citep{Herschel:1783lsr}, \footnote{\bf ``Now, if the proper motion of
the stars in general be once admitted, who can refuse to allow that
our sun, with all its planets and comets, that is, the solar sytsem,
is no less liable to such a general agitation as we find to obtain
among all the rest of the celestial bodies.  Admitting this for
granted, the greatest difficulty will be how to discern the proper
motion of the sun between so many other (and variously compounded)
motions of the stars.  This is an arduous task indeed, which we must
not hope to see accomplished in a little time; but we are not to be
discouraged from the attempt.  Let us, in all events, endeavour to lay
a good foundation for those who are to come after us.''}  and led to
the recognition that the Sun encounters interstellar clouds during its
journey through space \citep{Shapley:1921}.  Because both the Sun and
interstellar clouds move through space, the Doppler contribution of
solar motion to the heliocentric interstellar velocities must be
removed for comparisons between kinematically defined clouds and
spatially defined objects such as Loop I.}

For conversion to the LSR velocity frame, which traces the mean
velocity of nearby gravitationally relaxed stars around the galactic
center, we use the solar motion relative to the LSR derived by
\citet{SchonrichBinneyDehnen:2010}.  Using data from the Hipparcos
spacecraft \citep{Perrymanetal:1997}, the
\citet{SchonrichBinneyDehnen:2010} results for the U, V, and W solar
velocity components correspond to a solar velocity of V$=18.0\pm 0.9$
\kms\ toward $\ell=47.8^\circ \pm 2.9^\circ$, $B=23.8^\circ \pm
2.0^\circ$ (e.g. the solar apex motion).  {The direction determined by
\citet{Herschel:1783lsr}, based on the nearest and brightest stars,
was toward the star $\lambda$ Her that is located 5\deeg\ away from
this direction.}

The bulk motion through space of the nearby interstellar material
associated with the CLIC, $<30$ pc, been determined from the
velocities for interstellar optical and ultraviolet absorption lines,
as well as the flow of interstellar dust through the heliosphere.
Observations of interstellar absorption lines in 96 stars sampling the
nearby ISM have been used to determine the bulk flow of nearby ISM
through space for the assumption that the material flows as a
rigid-body \citep{FGW:2002}, yielding the heliocentric flow velocity
of $28.1 \pm 4.6$ \kms\ toward \glon,\glat=192.4\deeg, --11.6\deeg.
Uncertainties on the longitude or latitude were not originally
provided, so for the purposes of evaluating the motion of the CLIC
relative to the LSR, uncertainties of 2\deeg\ each are adopted for
\glon\ and \glat.

The heliocentric velocity of the CLIC, based on the bulk motion of the
CLIC in \citet{FGW:2002}, is given in Table \ref{tab:lsr}.  The LIC
LSR velocity in Table \ref{tab:lsr} is based on the IBEX \HeI\
velocity in \citet{Schwadron:2015isn}.

%%%%%%%%%%%%%%%%%%%%%%%%%%%%%%%%%%%%%%%%%%%%%%%%%%%%%%%%%%%%%%%%%%%%%%%%%%%%%%%%%%%%%%%%
\pagebreak
%\bibliography{ismf3-long,frisch,ibex2,gcr_snr}
%\bibliographystyle{apj}

%%%%%%%%%%%%%%%%%%%%%%%%%%%%%%%%%%%%%%%%%%%%%%%%%%%%%%%%%%%%%%%%%%%%%%%%%%%%%%%%%%%%%%%%
\begin{deluxetable}{lccc}
\tablecaption{Interstellar Magnetic Field Directions \label{tab:summary}}
\tablewidth{0pt} 
\tabletypesize{\small}
\footnotesize{\tiny}
\tablehead{\colhead{Magnetic Field} & \colhead{Merit\tablenotemark{(A)}} & \colhead{ISMF Direction\tablenotemark{(B)}} & \colhead{} \\
\colhead{} & \colhead{function} &  \colhead{$\ell,~b$ (deg.)} & \colhead{} }
\startdata 
Paper I &  Unweighted fit  &  $38,~23 ~( \pm 35) $ &   \\ 
Paper II & \meritfII   &  $47 \pm 15,~25 \pm 20 $ &  \\ 
All stars (\Ball) &  \meritfII & $16.3,~27.0 ~( \pm 15 )$ &  \\ 
Interstellar (\Bnofil, no filament stars) & \meritfII & $36.2,~49.0 ~( \pm 16 )$  &  \\
Filament (\Bfil, only filament stars)\tablenotemark{(C)}  & \meritfII\ & $359.3,~19.0$ ($\pm 10.2$)  &  \\
IBEX (\Bibex)\tablenotemark{(D)} &  & $34.8 \pm 4.3 ,~56.6 \pm 1.2$ &  \\
Angle between \Bnofil\ and \Vcliclsr\tablenotemark{(E)} & & 76.8 (+23.5, --27.6)  &  \\
Angle between \Bnofil\ and \Bibex & & $7.6 (+14.9, -7.6) $ &  \\
Angle between \Bibex\ and V$_\mathrm{LIC,LSR}$\tablenotemark{F} & & $ 96.9 \pm 8.5$ \\
 &  \\
\enddata 
\tablenotetext{(A)}{Eqn. \ref{eqn:meritfII}}
\tablenotetext{(B)}{The quantities $\ell,~b$ are the direction of the ISMF in galactic coordinates.}
\tablenotetext{(C)}{Direction of the ISMF traced the polarization filament, from \citet{Frisch:2015fil}}
\tablenotetext{(D)}{ISMF direction traced by the IBEX Ribbon, \Bibex, corresponding
to the weighted mean of the energy-dependent center of the IBEX Ribbon arc at \elon$=219.2 \pm 1.3 $,
\elat$=39.9  \pm 2.3 $ \citep{Funsten:2013}.}
\tablenotetext{(E)}{Based on an upwind direction for the LSR CLIC velocity vector of
$\ell=335.6 \pm 13.4,~b=-7.0 \pm 9.0$, $V = -17.3 \pm 4.9$ (Appendix \ref{app:lsr}).}
\tablenotetext{(F)}{See Appendix \ref{app:lsr} for the LSR velocity of the LIC.}
\end{deluxetable}

\begin{deluxetable}{lccccl} 
\tablecaption{Polarization data \label{tab:data} }
\tablewidth{0pt}  
\footnotesize{\tiny} 
\tablehead{ 
\colhead{Star} & \colhead{$\ell,b$ } & \colhead{Distance} & \colhead{$\theta_\mathrm{RA}$\tablenotemark{A}} & \colhead{Polarization\tablenotemark{B}} & \colhead{Source\tablenotemark{C}} \nl 
\colhead{} & \colhead{(deg)} & \colhead{(pc)} & \colhead{(deg)} & \colhead{($10^{-5}$)} & \colhead{} } 
\startdata  
 HIP98130 &   9, -27 & 19 & $131.0 \pm 22.5 $  & $   23. \pm    23. $ & LNA   \nl
 HIP11276 & 354, -61 & 28 & $104.0 \pm  8.7 $  & $   32. \pm    10. $ & LNA   \nl
  HIP2790 & 316, -76 & 29 & $157.0 \pm 12.1 $  & $   31. \pm    14. $ & LNA   \nl
 HIP10301 & 279, -59 & 29 & $ 31.0 \pm  2.1 $  & $  124. \pm     9. $ & LNA   \nl
 HIP11197 &  14, -61 & 26 & $128.0 \pm 29.5 $  & $   27. \pm    45. $ & LNA   \nl
 HIP95467 & 329, -28 & 26 & $118.0 \pm 17.7 $  & $   24. \pm    17. $ & LNA   \nl
 HIP90355 &  37,  10 & 27 & $111.0 \pm  9.2 $  & $  120. \pm    40. $ & LNA   \nl
 HD105330 & 292,  31 & 33 & $ 95.0 \pm 25.9 $  & $   45. \pm    57. $ & LNA   \nl
  HD78351 & 258,   9 & 39 & $ 30.0 \pm 10.4 $  & $   42. \pm    16. $ & LNA   \nl
 HD111232 & 303,  -5 & 30 & $ 92.0 \pm 32.4 $  & $   30. \pm    64. $ & LNA   \nl
 HD128674 & 317,   3 & 27 & $ 63.0 \pm 16.7 $  & $   73. \pm    48. $ & LNA   \nl
 HD177409 & 327, -27 & 35 & $133.0 \pm 15.4 $  & $   37. \pm    22. $ & LNA   \nl
 HD117939 & 312,  23 & 30 & $ 40.0 \pm 10.2 $  & $   78. \pm    29. $ & LNA   \nl
 HD125162 &  87,  64 & 30 & $ 25.0 \pm 15.0 $  & $  2.72 \pm   1.45 $ & KVA   \nl
 HD130109 & 355,  52 & 39 & $ 45.2 \pm  8.3 $  & $  5.46 \pm   1.60 $ & KVA   \nl
 HD132052 & 351,  46 & 28 & $ 45.7 \pm 31.6 $  & $  1.88 \pm   2.32 $ & KVA   \nl
 HD137391 &  60,  56 & 37 & $ 24.5 \pm 22.3 $  & $  1.42 \pm   1.17 $ & KVA   \nl
 HD175638 &  37,   0 & 40 & $ 85.0 \pm 27.3 $  & $  2.10 \pm   2.16 $ & KVA   \nl
 HD185395 &  82,  13 & 19 & $ 67.1 \pm  8.7 $  & $  3.27 \pm   1.00 $ & KVA   \nl
  HD19373 & 144,  -7 & 11 & $ 25.2 \pm  8.7 $  & $  4.04 \pm   1.24 $ & KVA   \nl
  HD91889 & 258,  38 & 25 & $ 64.6 \pm 19.1 $  & $  5.09 \pm   3.52 $ & KVA   \nl
 HD124850 & 337,  51 & 21 & $148.5 \pm 23.0 $  & $  1.38 \pm   1.17 $ & KVA   \nl
  HD14055 & 142, -25 & 36 & $  7.9 \pm  4.4 $  & $  4.54 \pm   0.69 $ & KVA   \nl
  HD15335 & 146, -28 & 31 & $ 21.4 \pm 14.9 $  & $  4.28 \pm   2.28 $ & KVA   \nl
  HD18256 & 160, -35 & 35 & $ 28.2 \pm 12.4 $  & $  1.71 \pm   0.75 $ & KVA   \nl
  HD18404 & 158, -33 & 32 & $ 27.0 \pm 14.1 $  & $  1.03 \pm   0.52 $ & KVA   \nl
 HD206901 &  78, -20 & 35 & $  6.9 \pm 36.9 $  & $  0.82 \pm   1.23 $ & KVA   \nl
 HD222603 &  90, -56 & 31 & $ 21.7 \pm 37.3 $  & $  1.17 \pm   1.78 $ & KVA   \nl
  HD25490 & 184, -33 & 40 & $ 26.9 \pm 19.2 $  & $  2.24 \pm   1.56 $ & KVA   \nl
  HD25570 & 182, -31 & 36 & $ 24.5 \pm 10.9 $  & $  2.59 \pm   1.00 $ & KVA   \nl
   HD8829 & 155, -73 & 36 & $ 17.8 \pm 29.7 $  & $  2.20 \pm   2.51 $ & KVA   \nl
  HD13555 & 147, -37 & 30 & $ 11.3 \pm 13.8 $  & $  2.45 \pm   1.20 $ & KVA   \nl
\enddata  
\tablenotetext{A}{Polarization position angles are presented in equatorial coordinates.}
\tablenotetext{B}{{The polarizations represent the fractional linear polarizations of the E-component
of the starlight}}.
\tablenotetext{C}{{The data source refers to the telescope where the data were
acquired.  The KVA data were acquired by A. Berdyugin and V. Piirola using BVR filters.
The LNA data were acquired by A. M. Magalhaes and D. B. Seriacopi using the V filter.} }
\end{deluxetable}

\begin{deluxetable}{lc}
\tablecaption{Summary of LSR vectors and directions } \label{tab:lsr}
\tablewidth{0pt} 
\tabletypesize{\small}
\footnotesize{\tiny}
\tablehead{\colhead{Quantity} & \colhead{LSR Vector Velocity} \\
\colhead{} & \colhead{$\ell$ (deg), $b$ (deg), V (\kms)} }
\startdata 
%% & $\ell$ (deg), $b$ (deg), V (\kms)
Vector velocity of solar apex motion\tablenotemark{A}    & $ 47.8 \pm 2.9,~23.8 \pm 2.0,~18.0 \pm 0.9$  \\
Vector velocity of bulk CLIC LSR motion\tablenotemark{B} &  $155.6 \pm 13.4, ~ 7.0 \pm 9.0,~ 17.3 \pm 4.9$  \\ 
Vector velocity of bulk LIC LSR motion\tablenotemark{C} &  $141.1 \pm 5.9, ~ 2.4 \pm 4.2,~ 17.2 \pm 1.9$   \\
\enddata
\tablenotetext{A}{Based on the U,V,W components of the solar apex motion in \citet{SchonrichBinneyDehnen:2010}.}
\tablenotetext{B}{Based on the heliocentric velocity vector of the CLIC derived in \citet{FGW:2002}, with the additional assumption
that uncertainties on the longitude and latitude of the vector are 2\deeg\ each.}
\tablenotetext{C}{This LSR velocity for the LIC is based on the  heliocentric velocity derived for the flow of interstellar \HeI\ through the
heliosphere that compared in situ IBEX-LO \HeI\ data with multivariate simulations of the particle trjacetories \citep{Schwadron:2015isn}.}
\end{deluxetable}

\newpage
\pagebreak

%%%%%%%%%%%%%%%%%%%%%%%%%%%%%%%%%%%%%%%%%%%%%%%%%%%%%%%%%%%%%%%%%%%%%%%%%%%%%%%%%%%%%%%%

\end{document}